\documentclass[final]{dmtcs-episciences}
\newcommand{\dmtcs}{}

\usepackage[bold]{complexity}
\usepackage[round]{natbib}

\usepackage[usenames,dvipsnames]{xcolor}

\newcommand{\chosenlanguage}{english}
\usepackage{ifthen}
\usepackage{multirow}
\usepackage{color}
\usepackage{colortbl}
\usepackage{array}
\usepackage[normalem]{ulem}
\usepackage[utf8]{inputenc}
\usepackage[T1]{fontenc}
\ifthenelse{\isundefined{\chosenlanguage}}{
    \input{french}
}{
    \usepackage[english]{babel}
\usepackage[ruled,noend,noline]{algorithm2e}
\ifthenelse{\isundefined{\llncs}}{
    \usepackage{amsthm}

\theoremstyle{plain}

\ifthenelse{\isundefined{\bpresentation}}{
    
}{}

\newtheorem{nthrm}{Theorem}

\newtheorem{nlemma}{Lemma}

\newtheorem{nobservation}{Observation}
\newtheorem{npb}{Problem}
\newtheorem{npbopt}{}%Optimization Problem}
\newtheorem{npbdec}{Decision Problem}

\newtheorem*{thrm*}{Theorem}
\newtheorem*{lemma*}{Lemma}
\newtheorem*{corol*}{Corollary}
\newtheorem*{trule*}{Rule}
\theoremstyle{definition}

\newtheorem{nprop}{Property}

\newtheorem*{df*}{Definition}
\newtheorem*{prop*}{Property}
\newtheorem*{conj*}{Conjecture}
\theoremstyle{remark}
\ifthenelse{\isundefined{\bpresentation}}{
    
}{}

\newtheorem*{note*}{Note}
\newtheorem*{pr*}{Proposition}
\newtheorem*{rmq*}{Remark}
\newtheorem*{not*}{Notation}
}{}

\newcommand{\pbinput}{Input}
\newcommand{\pboutput}{Output}
\newcommand{\pbparam}{Parameter}

}
\usepackage{eurosym}
\usepackage{xargs}
\usepackage{xstring}
\usepackage{graphicx}
\ifthenelse{\isundefined{\usesubfig}}
{
}{
    \usepackage{subfig}
}
\usepackage{float}
\usepackage{rotating}
\usepackage{moreverb}
\usepackage{pdfpages}
\usepackage{varwidth}
\ifthenelse{\isundefined{\bpresentation}}{
    \ifthenelse{\isundefined{\largepages}}{}{
        \usepackage[top=2.5cm, bottom=2.5cm, left=2.5cm, right=2.5cm]{geometry}
    }
    \usepackage{fancyhdr}
}{}
\ifthenelse{\isundefined{\bpresentation}}{
    \usepackage[reqno]{amsmath}
}{
    \newsavebox{\adjusted}
    \newlength{\adjustedlength}
    
}

\usepackage{amssymb}
\usepackage{bbm}
\usepackage{tikz}
\usepackage{pgf}
\usetikzlibrary{arrows,automata,shapes,fit,backgrounds}
\usetikzlibrary{petri}
\usetikzlibrary{trees}
\ifthenelse{\isundefined{\schedpackage}}{}{
    \usepackage{pgfgantt}
    \usepackage{scheduling}
}
\usepackage{appendix}
\renewcommand{\O}{\mathcal{O}}

\def\pbbodywidth{0.81}
\def\pbheadwidth{0.13}

\newcommand{\dfopt}[4][nolabel]{
    \noindent
    \begin{minipage}{\linewidth}
        ~\\
        \hrule
        \begin{npbopt}
            \ifthenelse{\equal{#1}{nolabel}}{}{\label{#1}}
            #2
            \\
            \vspace{0.2em}
            \\
            \begin{tabular}{p{\pbheadwidth\linewidth}p{\pbbodywidth\linewidth}}
                \textbf{\pbinput} & #3 \\[0.8em]
                \textbf{\pboutput} & #4 \\
            \end{tabular}
        \end{npbopt}
        \hrule
        ~\\
    \end{minipage}
}

\newcommand{\dfdec}[4][nolabel]{
    \noindent
    \begin{minipage}{\linewidth}
        ~\\
        \hrule
        \begin{npbdec}
            \ifthenelse{\equal{#1}{nolabel}}{}{\label{#1}}
            #2
            \\
            \vspace{0.2em}
            \\
            \begin{tabular}{p{0.18\linewidth}p{0.8\linewidth}}
                \textbf{\pbinput} & #3 \\[0.8em]
                \textbf{\pboutput} & #4 \\
            \end{tabular}
        \end{npbdec}
        \hrule
        ~\\
    \end{minipage}
}

\newcommand{\dfpb}[4][nolabel]{
    \noindent
    \begin{minipage}{\linewidth}
        ~\\
        \hrule
        \begin{npb}
            \ifthenelse{\equal{#1}{nolabel}}{}{\label{#1}}
            #2
            \\
            \vspace{0.2em}
            \\
            \begin{tabular}{p{0.18\linewidth}p{0.8\linewidth}}
                \textbf{\pbinput} & #3 \\[0.8em]
                \textbf{\pboutput} & #4 \\
            \end{tabular}
        \end{npb}
        \hrule
        ~\\
    \end{minipage}
}

\newcommand{\dfpbp}[5][nolabel]{
    \noindent
    \begin{minipage}{\linewidth}
        ~\\
        \hrule
        \begin{npb}
            \ifthenelse{\equal{#1}{nolabel}}{}{\label{#1}}
            #2
            \\
            \vspace{0.2em}
            \\
            \begin{tabular}{p{0.18\linewidth}p{0.8\linewidth}}
                \textbf{\pbinput} & #3 \\[0.8em]
                \textbf{\pbparam} & #5 \\[0.8em]
                \textbf{\pboutput} & #4 \\
            \end{tabular}
        \end{npb}
        \hrule
        ~\\
    \end{minipage}
}
\tikzstyle{edge}=[ -, draw=black ]
\tikzstyle{arrow}=[->, draw=black, >=latex ]

\tikzstyle{sized}=[
    minimum width=9mm
]

\tikzstyle{tvertex}=[
	node distance=15mm,
    inner sep = 1mm
]

\tikzstyle{tvert}=[
	node distance=5mm,
    inner sep=1mm
]

\tikzstyle{tmvertex}=[
    tvertex,
    sized
]

\tikzstyle{tmvert}=[
    tvert,
    sized
]

\tikzstyle{mvertex}=[
    tmvertex,
	circle,
	draw=black,
	fill=black!50
]

\tikzstyle{mvert}=[
    tmvert,
	circle,
	draw=black,
	fill=black!50
]

\tikzstyle{vertex}=[
    tvertex,
	circle,
	draw=black,
	fill=black!50
]

\tikzstyle{vert}=[
    tvert,
	circle,
	draw=black,
	fill=black!50
]

\tikzstyle{treedec}=[
	vertex,
	node distance=15mm,
	draw=red,
	fill=red!50
]

\tikzstyle{treeedge}=[
	edge,
	draw=red
]

\tikzstyle{giedge}=[
	edge,
	draw=blue
]

\tikzstyle{bounding}=[
	thick, %ultra
	densely dotted,
	fill=blue!10,
	draw=blue!70
]

\tikzstyle{patate}=[
    rounded corners=5pt,
    draw=black,
    fill=black!50,
    text width=12mm,
    text centered
]

\tikzstyle{patatoid}=[
    ellipse,
    node distance=10mm,
    draw=black
]
\newcounter{psubfig}

\newcommand{\ie}[0]{\textit{i.e.}\xspace}

\newcommand{\vset}[2][\vsetnot]{\ensuremath{#1^{#2}}}

\newcommand{\nset}[0]{\ensuremath{m}}
\newcommand{\nvec}[0]{\ensuremath{n}}
\newcommand{\ndie}[0]{\ensuremath{p}}

\newcommand{\cfunc}[2][\cfuncnot]{\ensuremath{#1(#2)}}

\newcommand{\mtuple}[1][\nset]{\ensuremath{#1}-tuple\xspace}
\newcommand{\mtuples}[1][\nset]{\ensuremath{#1}-tuples\xspace}

\ifthenelse{\isundefined{\dmtcs}}{

}{
\newtheorem{definition}{Definition}
}
\usepackage{enumerate}

\floatstyle{ruled}
\newfloat{problem}{th}{prob}
\floatname{problem}{Problem}

\newcommand{\chicol}{\ensuremath{\chi}-\textsc{Coloring}\xspace}
\newcommand{\probFPT}{\textsc{Multidimensional Binary Vector Assignment}\xspace}
\newcommand{\probFPTshort}{\textsc{bMVA}\xspace}

\newcommand{\eg}{\textit{e.g.\xspace}}
\newcommand{\abovep}{\ensuremath{\zeta_p}}
\newcommand{\aboveB}{\ensuremath{\zeta_{\mathcal{B}}}}
\newcommand{\B}{\ensuremath{\mathcal{B}}}
\newcommand{\oct}{\textsc{Odd Cycle Transversal}\xspace}
\newcommand{\octshort}{\textsc{OCT}\xspace}
\newcommand{\bipoct}{\textsc{bip-OCT}\xspace}
\newcommand{\dualprob}{\ensuremath{max \sum 1}-\probFPTshort}
\newcommand{\I}{\ensuremath{\mathcal{I}}}
\title[Multidimensional Binary Vector Assignment problem]{Multidimensional Binary Vector Assignment problem: standard, structural and above guarantee parameterizations}

\author[M. Bougeret, G. Duvillié, R. Giroudeau, R. Watrigant]%
{
	Marin Bougeret\affiliationmark{1} 
    \and Guillerme Duvilli\'e\affiliationmark{1}
    \and Rodolphe Giroudeau\affiliationmark{1} 
    \and Rémi Watrigant\affiliationmark{2}
}

\affiliation{
    LIRMM, Universit\'e Montpellier 2, France\\%
Hong Kong Polytechnic University. Computing department}

\received{2015-11-12}
\revised{2016-12-5}
\accepted{2017-10-3}
\publicationdetails{19}{2017}{4}{3}{1331}


\begin{document}
\maketitle

\newcommand{\cratio}{\ensuremath{r}}

\def\figscale{0.9}

% Array Measure
\def\arrayscale{0.4em}
\def\headwidth{0.14\linewidth}
\def\bodywidth{0.80\linewidth}

\begin{abstract}
    In this article we focus on the parameterized complexity of the Multidimensional Binary Vector
    Assignment problem (called \probFPTshort).  An input of this problem is defined by \nset{}
    disjoint sets \begin{math}V^1, V^2, \dots, V^m\end{math}, each composed of \nvec{} binary
    vectors of size \ndie{}. An output is a set of \nvec{} disjoint \mtuples of vectors, where each
    \mtuple is obtained by picking one vector from each set \begin{math}V^i\end{math}.  To each
    \mtuple we associate a \ndie{} dimensional vector by applying the bit-wise AND operation on the
    \nset{} vectors of the tuple.  The objective is to minimize the total number of zeros in these
    \nvec{} vectors.  \probFPTshort can be seen as a variant of multidimensional matching where
    hyperedges are implicitly locally encoded via labels attached to vertices, but was originally
    introduced in the context of integrated circuit manufacturing.

    We provide for this problem FPT algorithms and negative results (\begin{math}ETH\end{math}-based
    results, \W{2}-hardness and a kernel lower bound) according to several parameters: the standard
    parameter \begin{math}k\end{math} (\textit{i.e.} the total number of zeros), as well as two
    parameters above some guaranteed values.

    %A preliminary version of this work has been published in~\cite{bougeret2015multidimensional}.
\end{abstract}

\section{Introduction}

\subsection{Definition of the problem}

In this paper, we consider the parameterized version of the \probFPT{} problem (\probFPTshort). An
input of this problem is described by \nset{} sets \begin{math}V^1, V^2, \dots, V^m\end{math}, each
of these sets containing \nvec{} \ndie{}-dimensional binary vectors. We note \begin{math}V^i =
    \{v_1^i, \dots, v_n^i\}\end{math} for all \begin{math}i \in
    [m]\end{math}\footnote{\begin{math}[m]\end{math} stands for \begin{math}\left\{ 1, \dots,
    m\right\}\end{math}.}, and for all \begin{math}j \in [n]\end{math} and \begin{math}r \in
[p]\end{math}, we denote by \begin{math}v_j^i[r] \in \{0, 1\}\end{math} the
\begin{math}r^{th}\end{math} component of \begin{math}v_j^i\end{math}.

In order to define the output of the problem, we need to introduce the notion of stack. A stack
\begin{math}s = (v_1^s, v_2^s, \dots, v_m^s)\end{math} is an \mtuple of vectors such that
\begin{math}\forall i \in [m], v_i^s \in V^i\end{math}. The output of \probFPTshort is a set
\begin{math}S\end{math} of \nvec{} stacks such that for all \begin{math}i, j \in [m] \times
[n]\end{math}, \begin{math}v_j^i\end{math} belongs to only one stack (in that case, the stacks are
said \textit{disjoint}). An example of an instance together with a solution is depicted in
Figure~\ref{fig:exemple1}.

We are now ready to define the objective function. We define the operator
\begin{math}\wedge\end{math} that, given two \ndie{}-dimensional vectors \begin{math}u\end{math} and
\begin{math}v\end{math}, computes the vector \begin{math}w = \left( u[1] \wedge v[1], u[2] \wedge
v[2], \dots, u[p] \wedge v[p]\right)\end{math}. We associate to any stack \begin{math}s\end{math} a
unique vector \begin{math}v_s = \bigwedge_{i \in [m]} v_{i}^s\end{math}.

We define the cost of a binary vector \begin{math}v\end{math} as the number of zeros in it. More
formally, if \begin{math}v\end{math} is \ndie{}-dimensional, \begin{math}c(v) = p - \sum_{r \in [p]}
    v[r]\end{math}. We extend this definition to a set of stacks \begin{math}S = \{s_1, \dots,
s_n\}\end{math} as follows: \begin{math}c(S) = \sum_{j \in [n]} c(v_{s_j})\end{math}.  Finally, the
objective of \probFPTshort is to obtain a set \begin{math}S\end{math} of \nvec{} disjoint stacks
while minimizing \begin{math}c(S)\end{math}. In the decision version of the problem, we are given an
integer \begin{math}k\end{math}, and we ask whether there exists a solution \begin{math}S\end{math}
of cost at most \begin{math}k\end{math}. The problem is thus defined formally as follows:

\begin{problem}[H]
    \begin{minipage}{\linewidth}
        \begin{tabular}{m{\headwidth}m{\bodywidth}}
            \textbf{Input:} & \nset{} sets of \nvec{} binary \ndie{}-dimensional vectors, an integer
            \begin{math}k\end{math} \\
            \textbf{Question:} & Is there a set \begin{math}S\end{math} of \nvec{} disjoint stacks
            such that \begin{math}c(S) \le k\end{math} ? \\
        \end{tabular}
    \end{minipage}
    \caption{\probFPT (\probFPTshort)}
    \label{pb:wwi_parameterized}
\end{problem}

\begin{figure}
    \begin{center}
        \begin{tikzpicture}[wafer/.style={rectangle, draw}, every node/.style={transform shape},scale=\figscale]

            \def\yscale{0.6}
            \node at (0, \yscale *5) {\vset{1}};
            \node at (2, \yscale *5) {\vset{2}};
            \node at (4, \yscale *5) {\vset{3}};
            \node at (7, \yscale *5) {S};

            \node[wafer] (v11) at (0, \yscale* 4) {\begin{math}001101\end{math}};
            \node[wafer] (v12) at (0, \yscale* 3) {\begin{math}110111\end{math}};
            \node[wafer] (v13) at (0, \yscale* 2) {\begin{math}011101\end{math}};
            \node[wafer] (v14) at (0, \yscale* 1) {\begin{math}111101\end{math}};

            \node[wafer] (v21) at (2, \yscale* 4) {\begin{math}110010\end{math}};
            \node[wafer] (v22) at (2, \yscale* 3) {\begin{math}010101\end{math}};
            \node[wafer] (v23) at (2, \yscale* 2) {\begin{math}110011\end{math}};
            \node[wafer] (v24) at (2, \yscale* 1) {\begin{math}010101\end{math}};

            \node[wafer] (v31) at (4, \yscale* 4) {\begin{math}110110\end{math}};
            \node[wafer] (v32) at (4, \yscale* 3) {\begin{math}010110\end{math}};
            \node[wafer] (v33) at (4, \yscale* 2) {\begin{math}010011\end{math}};
            \node[wafer] (v34) at (4, \yscale* 1) {\begin{math}001111\end{math}};

            \draw[thick, dashed] (v12) to (v21) to (v31); 
            \draw[thick] (v11) to (v23) to (v32); 
            \draw[thick, dashdotted] (v13) to (v24) to (v33); 
            \draw[thick, dotted] (v14) to (v22) to (v34); 

            % \node[wafer, color=orange!80]  (vs1) at (8, 4) {\begin{math}(110111, 110010,
            % 110110)\end{math}};
            % \node[wafer, color=red!80]     (vs2) at (8, 3) {\begin{math}(001101, 110011,
            % 010110)\end{math}};
            % \node[wafer, color=blue!80]    (vs3) at (8, 2) {\begin{math}(011101, 010101,
            % 010011)\end{math}};
            % \node[wafer, color=magenta!80] (vs4) at (8, 1) {\begin{math}(111101, 010101,
            % 001111)\end{math}};
%             \node[wafer, color=orange!80]  (vs1) at (7, 4) {\begin{math}110010\end{math}};
%             \node[wafer, color=red!80]     (vs2) at (7, 3) {\begin{math}000000\end{math}};
%             \node[wafer, color=blue!80]    (vs3) at (7, 2) {\begin{math}010001\end{math}};
%             \node[wafer, color=magenta!80] (vs4) at (7, 1) {\begin{math}000101\end{math}};
%             \node[color=orange!80]               at (8, 4) {\begin{math}v_{s_{1}}\end{math}};
%             \node[color=red!80]                  at (8, 3) {\begin{math}v_{s_{2}}\end{math}};
%             \node[color=blue!80]                 at (8, 2) {\begin{math}v_{s_{3}}\end{math}};
%             \node[color=magenta!80]              at (8, 1) {\begin{math}v_{s_{4}}\end{math}};
% 
            \node[wafer]  (vs1) at (7, \yscale*4) {\begin{math}110010\end{math}};
            \node[wafer]  (vs2) at (7, \yscale*3) {\begin{math}000000\end{math}};
            \node[wafer]  (vs3) at (7, \yscale*2) {\begin{math}010001\end{math}};
            \node[wafer]  (vs4) at (7, \yscale*1) {\begin{math}000101\end{math}};
            \node               at (8, \yscale*4) {\begin{math}v_{s_{1}}\end{math}};
            \node               at (8, \yscale*3) {\begin{math}v_{s_{2}}\end{math}};
            \node               at (8, \yscale*2) {\begin{math}v_{s_{3}}\end{math}};
            \node               at (8, \yscale*1) {\begin{math}v_{s_{4}}\end{math}};

            \node               at (10, \yscale*4) {\begin{math}\cfunc{v_{s_{1}}} = 3\end{math}};
            \node               at (10, \yscale*3) {\begin{math}\cfunc{v_{s_{2}}} = 6\end{math}};
            \node               at (10, \yscale*2) {\begin{math}\cfunc{v_{s_{3}}} = 4\end{math}};
            \node               at (10, \yscale*1) {\begin{math}\cfunc{v_{s_{4}}} = 4\end{math}};

            \draw[thick, dashed] (v31) to (vs1);
            \draw[thick, dotted] (v34) to (vs4);            
            \draw[thick, dashdotted] (v33) to (vs3);
            \draw[thick] (v32) to (vs2);

            \draw[thick, dashed]     (12, \yscale * 4) to node[right] {\begin{math}\quad s_1\end{math}} (12.4, \yscale*4);
            \draw[thick, dotted]     (12, \yscale * 3) to node[right] {\begin{math}\quad s_2\end{math}} (12.4, \yscale*3);
            \draw[thick, dashdotted] (12, \yscale * 2) to node[right] {\begin{math}\quad s_3\end{math}} (12.4, \yscale*2);
            \draw[thick]             (12, \yscale * 1) to node[right] {\begin{math}\quad s_4\end{math}} (12.4, \yscale*1);

        \end{tikzpicture}
%        \onslide<6->{
%        \begin{math}\cfunc{S} = \cfunc{110000} + \cfunc{000000} + \cfunc{010001} + \cfunc{000101} = 2 + 0 + 2 +
%        2 = 6\end{math}
%    }
    \end{center}
    \caption{Example of \probFPTshort{} instance with $\nset = 3, \nvec = 4, \ndie = 6$ and of a
    feasible solution $S$ of cost $c(S) = 17$.}
    \label{fig:exemple1}
\end{figure}

In order to avoid heavy notations throughout the paper, we will denote an instance of \probFPTshort
only by \begin{math}\mathcal{I}[m, n, p, k]\end{math}, the notations of the sets and vectors being
implicitly given as previously.

\subsection{Application and related work}

\probFPTshort can be seen as a variant of multidimensional matching where hyperedges are implicitly
locally encoded via labels attached to vertices.  However, this kind of problem was originally
introduced by~\cite{reda2009maximizing} in the context of semiconductor industry as the ``yield
maximization problem in wafer-to-wafer 3-D integration technology''. In this context, each vector
\begin{math}v^i_j\end{math} represents a wafer, which is seen as a string of bad dies (0) and good
dies (1). Integrating two wafers corresponds to superimposing the two corresponding strings. In this
operation, a position in the merged string is ``good'' when the two corresponding dies are good, and
is ``bad'' otherwise.  The objective of Wafer-to-Wafer Integration is to form \nvec{} stacks, while
maximizing their overall quality, or equivalently, minimizing the number of errors (depending on the
objective function). In the following, we will denote by \dualprob the dual version of \probFPTshort
where given the same input and ouput, the objective is to maximize \begin{math}np-c(S)\end{math},
the total number of ones.

The results obtained so far concerning these problems mainly concern their approximability. It is
proved in \cite{dokka2013approximation} that when $m=3$, \probFPTshort is
\begin{math}\NP{}\end{math}-hard but admits a \begin{math}\frac{4}{3}\end{math}-approximation.  We
can also mention~\cite{dokka2014multi} which
provides a \begin{math}f(m)\end{math}-approximation for general \nset{}, and an
\begin{math}\APX{}\end{math}-hardness for \begin{math}m=3\end{math}.  The main related article
is~\cite{bougeret2016complexity} where it is proved that \dualprob has no
\begin{math}O(p^{1-\epsilon})\end{math} nor \begin{math}O(m^{1-\epsilon})\end{math}-approximation
for any \begin{math}\epsilon > 0\end{math} unless \begin{math}\P{}=\NP{}\end{math} (even when
\begin{math}n=2\end{math}), but admits a \begin{math}\frac{p}{c}\end{math}-approximation algorithm
for any constant \begin{math}c \in \mathbb{N}\end{math}, and is \begin{math}\FPT{}\end{math} when
parameterized by \ndie{} (which also holds for \probFPTshort).  Notice that one of the reductions
provided by~\cite{bougeret2016complexity} is a parameter-preserving reduction from the \textsc{Clique}
problem to \dualprob, immediately proving \begin{math}\W{1}\end{math}-hardness for \dualprob when
parameterized by the objective function.  This is why our motivation in this paper is to consider
the parameterized complexity of \probFPTshort. As we will see in the next section, we provide an
analysis for several parameters related to this problem.\\

For formal definitions and detailed concepts on Fixed-Parameter Tractability, we refer to the
monograph of~\cite{downey2013fundamentals}. Moreover, in order to define lower bounds on the
running time of parameterized algorithms, we will rely on the \textit{Exponential Time Hypothesis}
(\begin{math}ETH\end{math}) of~\cite{impagliazzo2001problems}, stating that
\textsc{\begin{math}3\end{math}-Sat} cannot be solved in \begin{math}O^*(2^{o(n)})\end{math} where
\nvec{} is the number of variables (\begin{math}\O^*(.)\end{math} hides polynomial terms). For more
results about lower bounds obtained under \begin{math}ETH\end{math}, we refer the reader to the
survey of~\cite{lokshtanov2013lower}.

\subsection{Parameterizations}
\label{sec:param}

One of the main purposes of Fixed-Parameter Tractability is to obtain efficient algorithms when the
considered parameter is small in practice. When dealing with the decision version of an optimization
problem, the most natural parameter is perhaps the value of the desired solution
(\eg~\begin{math}k\end{math} for \probFPTshort).  Such a parameter is often referred to in the
literature as the ``standard parameter'' of the problem. In some cases, this parameter might not be
very interesting, either because it usually takes high values in practice, or because
\begin{math}\FPT{}\end{math} algorithms with respect to this parameterization are trivial to find.
When this happens, it is possible to obtain more interesting results by subtracting to the objective
function a known lower bound of it.  For instance, if one can prove that any solution of a given
minimization problem is of cost at least \B, then one can ask for a solution
of cost \begin{math}\mathcal{B}+c\end{math} and parameterize by \begin{math}c\end{math}.  This
idea, called ``above guarantee parameterization'' was introduced by~\cite{mahajan2009parameterizing} and first
applied to \textsc{Max Sat} and \textsc{Max Cut} problems. It then became a fruitful line of
research with similar results obtained for many other problems (among others,
see~\cite{cygan2013multiway,gutin2007linear,GY12,mahajan2009parameterizing}).

In this paper, we analyze the parameterized complexity of \probFPTshort using three types of
parameters. The first one is the standard parameter \begin{math}k\end{math}: the number of zeros to minimize in the
        optimization version of the problem. Then, three natural structural parameters: \nset{}, the number of sets of the input, \nvec{},
        the number of vectors in each set, and \ndie{}, the size of each vector. The last two parameters are above guarantee parameters.

As we said previously, we already proved in~\cite{bougeret2016complexity} that \probFPTshort is
\begin{math}\FPT{}\end{math} parameterized by \ndie{}. As we will notice in Lemma~\ref{lemma:boundp}
that we can obtain \begin{math}p \le k\end{math} after a polynomial pre-processing step, this
implies that \probFPTshort is also \begin{math}\FPT{}\end{math} with its standard parameter. Our
idea here is to use this previous inequality in order to obtain smaller parameters. Thus, we define
our first above guarantee parameter \begin{math}\abovep = k-p\end{math}.

Finally, in order to define our last parameter, we first need to describe the corresponding lower
bound \B, that will represent the maximum, over all sets of vectors, of the
total number of zeros for each set. More formally, we define \begin{math}\B = \max_{i \in [m]}
c(V^i)\end{math} where \begin{math}c(V^i) = \sum_{j=1}^n c(v_j^i)\end{math}. Since we perform a
bit-wise AND over each \mtuple, it is easily seen that any solution will be of cost at least
\begin{math}\B\end{math}. Thus, we define our last parameter \begin{math}\aboveB = k-\B\end{math}.

\subsection{Our results}

In the next section, we present some pre-processing rules leading to a kernel of size
\begin{math}O(k^2m)\end{math}, and prove that even when \begin{math}m=3\end{math} we cannot improve
it to \begin{math}p^{O(1)}\end{math} unless \begin{math}\NP{} \subseteq \coNP/\poly\end{math}
(remember that \probFPTshort is known to be \FPT{} when parameterized by \ndie{}).
Section~\ref{sec:b} is mainly focused on results associated with parameter
\begin{math}\aboveB\end{math}: we prove that \probFPTshort can be solved in
\begin{math}O^*(4^{\aboveB \log(n)})\end{math}, while it is \begin{math}\W{2}\end{math}-hard when
parameterized by \begin{math}\aboveB\end{math} only, and cannot be solved in
\begin{math}O^*(2^{o(\aboveB) \log(n)})\end{math} nor in \begin{math}O^*(2^{\aboveB
o(\log(n))})\end{math} assuming \begin{math}ETH\end{math}.  In Section~\ref{sec:p}, we focus on the
parameterization by \begin{math}\abovep\end{math}: we show that when \begin{math}n=2\end{math}, the
problem can be solved in single exponential time with this parameter, but is not in
\begin{math}\XP{}\end{math} for any fixed \begin{math}n \geq 3\end{math} (unless \begin{math}\P{} =
    \NP{}\end{math}). The reduction we use also shows that for fixed \begin{math}n \in
\mathbb{N}\end{math}, the problem cannot be solved in \begin{math}2^{o(k)}\end{math}  (and thus in
\begin{math}2^{o(\aboveB)}\end{math}) unless \begin{math}ETH\end{math} fails, which matches the
upper bound obtained in Section~\ref{sec:b}.  A summary of our results is depicted in the following
table.\\

\begin{tabular}{|m{0.25\linewidth}|m{0.65\linewidth}|} \hline
    \textbf{Positive results} & \textbf{Negative results} \\
    \hline
    \begin{math}O(k^2m)\end{math} kernel (Thm.~\ref{thm:kernelkm}) & no
    \begin{math}p^{O(1)}\end{math} kernel for $\nset = 3$ unless \begin{math}\NP \subseteq \coNP/\poly\end{math} (Thm.~\ref{thm:kernellowerboundp}) \\
    \hline
    \multirow{3}{\linewidth}{\begin{math}\O^*(4^{\aboveB \log(n)})\end{math} algorithm
    (Thm.~\ref{thm:fptaboveB})} & \begin{math}\W{2}\end{math}-hard for \begin{math}\aboveB\end{math} only (Thm.~\ref{thm:whardc})\\
    & no \begin{math}2^{o(\aboveB) \log(n)}\end{math} nor \begin{math}2^{\aboveB
    o(\log(n))}\end{math} under \begin{math}ETH\end{math} (Thm.~\ref{thm:lowerboundeth}) \\
    & no \begin{math}2^{o(k)}\end{math} for fixed \nvec{} under \begin{math}ETH\end{math} (Thm.~\ref{thm:reducfromcoloring}) \\
    \hline
    \begin{math}O^*(d^{\abovep})\end{math} algorithm &
    \multirow{2}{*}{\begin{math}\NP\end{math}-hard for \begin{math}\abovep=0\end{math} and fixed
    \begin{math}n \geq 3\end{math} (Thm.~\ref{thm:reducfromcoloring})}\\
    for \begin{math}n=2\end{math} (Thm.~\ref{thm:fptabovep})  & \\
    \hline
\end{tabular}
~\\

This article is the complete version of~\cite{bougeret2015multidimensional} where some proofs were
omitted due to space limitations.

\section{First remarks and kernels}\label{sec:preprocess}

\begin{figure}[h]
    \begin{center}
        \begin{tikzpicture}[wafer/.style={rectangle, draw}, every node/.style={transform shape},
            scale=\figscale]
            \def\yscale{0.6}
            \node at (0, \yscale *5) {\begin{math}\vset{1}\end{math}};
            \node at (2, \yscale *5) {\begin{math}\vset{2}\end{math}};
            \node at (4, \yscale *5) {\begin{math}\vset{3}\end{math}};
            \node at (7, \yscale *5) {\begin{math}Opt\end{math}};

            \node[wafer] (v11) at (0, \yscale* 4) {\begin{math}111\end{math}};
            \node[wafer] (v12) at (0, \yscale* 3) {\begin{math}101\end{math}};
            \node[wafer] (v13) at (0, \yscale* 2) {\begin{math}011\end{math}};

            \node[wafer] (v21) at (2, \yscale* 4) {\begin{math}111\end{math}};
            \node[wafer] (v22) at (2, \yscale* 3) {\begin{math}101\end{math}};
            \node[wafer] (v23) at (2, \yscale* 2) {\begin{math}110\end{math}};

            \node[wafer] (v31) at (4, \yscale* 4) {\begin{math}111\end{math}};
            \node[wafer] (v32) at (4, \yscale* 3) {\begin{math}011\end{math}};
            \node[wafer] (v33) at (4, \yscale* 2) {\begin{math}110\end{math}};

            \draw[thick, dashed] (v11) to (v23) to (v33); 
            \draw[thick] (v12) to (v22) to (v31); 
            \draw[thick, dotted] (v13) to (v21) to (v32); 

            \node[wafer]  (vs1) at (7, \yscale*4) {\begin{math}101\end{math}};
            \node[wafer]  (vs2) at (7, \yscale*3) {\begin{math}011\end{math}};
            \node[wafer]  (vs3) at (7, \yscale*2) {\begin{math}110\end{math}};
            \node               at (8, \yscale*4) {\begin{math}v_{s_{1}}\end{math}};
            \node               at (8, \yscale*3) {\begin{math}v_{s_{2}}\end{math}};
            \node               at (8, \yscale*2) {\begin{math}v_{s_{3}}\end{math}};

            \draw[thick, dashed] (v32) to (vs2);
            \draw[thick, dotted] (v33) to (vs3);            
            \draw[thick] (v31) to (vs1);

            \node               at (10, \yscale*4) {\begin{math}\cfunc{v_{s_{1}}} = 1\end{math}};
            \node               at (10, \yscale*3) {\begin{math}\cfunc{v_{s_{2}}} = 1\end{math}};
            \node               at (10, \yscale*2) {\begin{math}\cfunc{v_{s_{3}}} = 1\end{math}};
        \end{tikzpicture}
    \end{center}
    \caption{Example of \probFPTshort instance such that no optimal solution creates a
    $1$-stack.}
    \label{fig:stack1counterexample}
\end{figure}
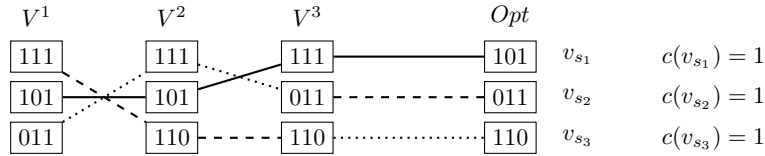

Let us start with two simple lemmas allowing us to bound the size of the input.  Notice first that
it is not always safe to create a \begin{math}1\end{math}-stack (\ie{} a stack with ones on every
    component) when possible.  Indeed, in instance \begin{math}V^1 = \left\{ \langle 111 \rangle,
        \langle 101 \rangle, \langle 011 \rangle \right\}\end{math}, \begin{math}V^2 = \left\{
    \langle 111 \rangle, \langle 101 \rangle, \langle 110 \rangle \right\}\end{math}, \begin{math}V^3 =
\left\{ \langle 111 \rangle, \langle 011 \rangle, \langle 110 \rangle \right\}\end{math}, depicted
in Figure~\ref{fig:stack1counterexample}, no optimal solution creates a
\begin{math}1\end{math}-stack. However, as we will see in Lemma~\ref{lemma:boundn}, creating
1-stacks becomes safe if \begin{math}n > k\end{math}.

\begin{nlemma}\label{lemma:boundn}
    There exists a polynomial algorithm which, given any instance \\\begin{math}\mathcal{I}[m, n, p,
    k]\end{math} of \probFPTshort, either detects that \begin{math}\mathcal{I}\end{math} is a
    negative instance, or outputs an equivalent instance \begin{math}\mathcal{I'}[m, n', p,
    k]\end{math} such that \begin{math}n' \le k\end{math}.
\end{nlemma}

\begin{proof}
    Let \begin{math}\mathcal{I}[m, n, p, k]\end{math} be an instance of \probFPTshort, and suppose
    that \begin{math}n > k\end{math}.  Let us write a polynomial pre-processing rule that either
    detects that \begin{math}\mathcal{I}\end{math} is a no instance, or compute an equivalent
    instance \begin{math}\mathcal{I}''[m,n'',p,k]\end{math} with \begin{math}n'' = n-1\end{math}.

    Notice first that there exists at least a 1-vector in every set \begin{math}V^i\end{math}. If
    not, \begin{math}\mathcal{I}\end{math} is a no instance as any solution would be of cost at
    least \begin{math}n>k\end{math}.  It is now safe to create a 1-stack, obtaining a remaining
    instance \begin{math}\mathcal{I}''\end{math} with \begin{math}n''=n-1\end{math}.  Indeed, if
    \begin{math}\mathcal{I}\end{math} is a yes instance, then there must exist at least one 1-stack
    in the solution (otherwise the cost would be at least \begin{math}n > k\end{math}), and thus the
    remaining instance is also a yes instance.  As the converse is trivially true, the rule is safe.
    Applying it at most \begin{math}n-k\end{math} times finally leads to the desired upper bound.
\end{proof}

%Using similar arguments, we can prove the following lemma:
%\begin{proof}
%    Let \begin{math}\mathcal{I}[m, n, p, k]\end{math} be an instance of the problem, and suppose
%    that there exists \begin{math}r
%    \in [p]\end{math} such that for all \begin{math}(i, j) \in [m] \times [n]\end{math} we have
%    \begin{math}v_j^i[r]=1\end{math}. In other words, the
%    \begin{math}r^{th}\end{math} component of all vectors of all sets is a \begin{math}1\end{math}. In this case, it is clear that all
%    vectors of any set of \nvec{} stacks obtained from \begin{math}\mathcal{I}\end{math} will also
%    contain a \begin{math}1\end{math} at the \begin{math}r^{th}\end{math}
%    component. Hence, we can modify \begin{math}\mathcal{I}[m, n, p, k]\end{math} into
%    \begin{math}\mathcal{I}'[m, n, p', k]\end{math} with \begin{math}p' < p\end{math} by dropping
%    all such components for all vectors. It is clear that this rule is safe since the cost of any
%    solution remains unchanged, and it can be applied in polynomial time. After applying this rule,
%    for all \begin{math}r \in [p']\end{math} there exists \begin{math}(i, j) \in [m] \times
%    [n]\end{math} such that \begin{math}v_j^i[r]=0\end{math}.
%This immediately implies that the cost of any solution is at least \begin{math}p'\end{math}.
%\qed
%\end{proof}

\begin{nlemma}\label{lemma:boundp}
    There exists a polynomial algorithm which, given any instance \\\begin{math}\mathcal{I}[m, n, p,
    k]\end{math} of \probFPTshort, either detects that \begin{math}\mathcal{I}\end{math} is a
    negative instance, or outputs an equivalent instance \begin{math}\mathcal{I'}[m, n, p',
    k]\end{math} such that \begin{math}p' \le k\end{math}.
\end{nlemma}

\begin{proof}
    Let \begin{math}\mathcal{I}[m, n, p, k]\end{math} be an instance of the problem, and suppose
    that there exists \begin{math}r \in [p]\end{math} such that for all \begin{math}(i, j) \in [m]
    \times [n]\end{math} we have \begin{math}v_j^i[r]=1\end{math}. In other words, the
    \begin{math}r^{th}\end{math} component of all vectors of all sets is a \begin{math}1\end{math}.
    In this case, it is clear that all vectors of any set of \nvec{} stacks obtained from
    \begin{math}\mathcal{I}\end{math} will also contain a \begin{math}1\end{math} at the
    \begin{math}r^{th}\end{math} component. Hence, we can modify \begin{math}\mathcal{I}[m, n, p,
        k]\end{math} into \begin{math}\mathcal{I}'[m, n, p', k]\end{math} with \begin{math}p' <
    p\end{math} by dropping all such components for all vectors. It is clear that this rule
    is safe since the cost of any solution remains unchanged, and it can be applied in
    polynomial time.  After applying this rule, for all \begin{math}r \in [p']\end{math}
    there exists \begin{math}(i, j) \in [m] \times [n]\end{math} such that
    \begin{math}v_j^i[r]=0\end{math}.  This immediately implies that the cost of any
    solution is at least \begin{math}p'\end{math}, and thus if \begin{math}p' > k\end{math}
    the algorithm detected a no instance.
\end{proof}

Given the two previous lemmas, we can suppose from now on that for any instance of \probFPTshort we
have \begin{math}n \le k\end{math} and \begin{math}p \le k\end{math}.  This immediately implies a
polynomial kernel parameterized by \begin{math}k\end{math} and \nset{}.

\begin{nthrm}\label{thm:kernelkm}
    \probFPTshort admits a kernel with \begin{math}O(k^2m)\end{math} bits.
\end{nthrm}

Let us now turn to the main result of this section.  To complement Theorem~\ref{thm:kernelkm}, we
show that even when \begin{math}m=3\end{math}, we cannot obtain a polynomial kernel with the smaller
parameter \ndie{} under some classical complexity assumptions (notice however that the existence of
a polynomial kernel in \begin{math}k\end{math} only is still open when \begin{math}m\end{math} is not fixed). Notice also that as \probFPTshort
was known to be \FPT{} when parameterized by \ndie{}, proved in~\cite{bougeret2016complexity}, it was a natural
question to ask for a polynomial kernel.

In order to establish kernel lower bounds, we use the concept of AND-cross-composition
of~\cite{bodlaender2014kernelization}, together with the recently proved AND-conjecture of~\cite{drucker2015new}. In the
following, a parameterized problem is a subset of \begin{math}\Sigma^* \times
\mathbb{N}\end{math}, where \begin{math}\Sigma^*\end{math} is the set of words over some finite
alphabet \begin{math}\Sigma\end{math}.

\begin{definition}[Polynomial equivalence relation according to~\cite{bodlaender2014kernelization}]
    An equivalence relation \begin{math}\mathcal{R}\end{math} on \begin{math}\Sigma^*\end{math} is
    called a polynomial equivalence relation if the two following conditions hold:
    \begin{itemize}
	    \item There is an algorithm that given two strings \begin{math}x, y \in \Sigma^*\end{math},
            decides whether \begin{math}x\end{math}
            and \begin{math}y\end{math} belong to the same equivalence class in
            \begin{math}(|x|+|y|)^{O(1)}\end{math} time.
	    \item For any finite set \begin{math}S \subseteq \Sigma^*\end{math}, the equivalence
            relation \begin{math}\mathcal{R}\end{math}
            partitions the elements of \begin{math}S\end{math} into at most \begin{math}(max_{x \in
            S} |x|)^{O(1)}\end{math} classes.
    \end{itemize}
\end{definition}

\begin{definition}[AND-cross-composition according to~\cite{bodlaender2014kernelization}]\label{def:andcross}
    Let \begin{math}L \subseteq \Sigma^*\end{math} be a set and let \begin{math}Q \subseteq \Sigma^*
    \times \mathbb{N}\end{math} be a parameterized problem. We say that \begin{math}L\end{math}
    AND-cross-composes into \begin{math}Q\end{math} if there is a polynomial equivalence relation
    \begin{math}\mathcal{R}\end{math} and an algorithm which, given \begin{math}t\end{math} strings
    belonging to the same equivalence class of \begin{math}\mathcal{R}\end{math}, computes an
    instance \begin{math}(x^*, k^*) \in \Sigma^* \times \mathbb{N}\end{math} in time polynomial in
    \begin{math}\sum_{i=1}^t |x_i|\end{math} such that:

    \begin{itemize}
	    \item \begin{math}(x^*, k^*) \in Q \Leftrightarrow x_i \in L\end{math} for all \begin{math}i
            \in \{1, \dots, t\}\end{math}
	    \item \begin{math}k^*\end{math} is bounded by a polynomial in \begin{math}max_{i=1}^t |x_i|
            + \log t\end{math}
    \end{itemize}
\end{definition}

\begin{nthrm}[\cite{drucker2015new}]\label{thm:andcross}
    If some set \begin{math}L \subseteq \Sigma^*\end{math} is \begin{math}\NP\end{math}-hard and
    \begin{math}L\end{math} AND-cross-composes into a parameterized problem \begin{math}Q\end{math},
    then there is no polynomial kernel for \begin{math}Q\end{math} unless \begin{math}\NP \subseteq
    \coNP/\poly\end{math}.
\end{nthrm}

\begin{nthrm}\label{thm:kernellowerboundp}
    Even for \begin{math}m=3\end{math}, \probFPTshort parameterized by \ndie{} does not admit a
    polynomial kernel unless \begin{math}\NP \subseteq \coNP/\poly\end{math}.
\end{nthrm}

\begin{proof}
    The proof is an AND-cross-composition inspired by the \begin{math}\NP\end{math}-hardness
    reduction for \probFPTshort provided by~\cite{dokka2013approximation}.  More precisely, we
    cross-compose from a sequence of instances of \textsc{\begin{math}3\end{math}-dimensional
    perfect matching}. According to Definition~\ref{def:andcross} and Theorem~\ref{thm:andcross},
    this will imply the desired result.  \textsc{\begin{math}3\end{math}-dimensional perfect
    matching} is formally defined as follows:

    \begin{problem}[H]
        \begin{minipage}{\linewidth}
            \begin{tabular}{m{\headwidth}m{\bodywidth}}
                \textbf{Input:} & Three sets \begin{math}X\end{math}, \begin{math}Y\end{math} and
                \begin{math}Z\end{math} of size \nvec{}, a set of hyperedges \begin{math}S \subseteq
                X \times Y \times Z\end{math} \\[\arrayscale]
                \textbf{Question:} & Does there exist a subset \begin{math}S' \subseteq S\end{math} such that: \\
                &		\begin{itemize}
			        \item for all \begin{math}e, e' \in S'\end{math} with \begin{math}e=(x, y,
                        z)\end{math} and \begin{math}e'=(x', y', z')\end{math}, we have
                        \begin{math}x \neq x'\end{math}, \begin{math}y \neq y'\end{math} and
                        \begin{math}z \neq z'\end{math} (that is, \begin{math}S'\end{math} is a matching)
			        \item \begin{math}|S'| = n\end{math} (that is, \begin{math}S'\end{math} is perfect)
		        \end{itemize}
            \end{tabular}
        \end{minipage}
        \caption{$3$-Dimensional Perfect Matching}
        \label{3DM}
    \end{problem}

    Let \begin{math}(X_1, Y_1, Z_1, S_1), \dots, (X_t, Y_t, Z_t, S_t)\end{math} be a sequence of
    \begin{math}t\end{math} equivalent instances of \begin{math}3\end{math}-DPM, with respect to the
    following polynomial equivalence relation: \begin{math}(X, Y, Z, S)\end{math} and
    \begin{math}(X', Y', Z', S')\end{math} are equivalent if \begin{math}|X| = |X'|\end{math} (and
    thus \begin{math}|Y|=|Y'|=|Z|=|Z'|=|X|\end{math}), and \begin{math}|S| = |S'|\end{math}. In the
    following we denote by \nvec{} the cardinality of the sets \begin{math}X_i\end{math} (and
    equivalently the sets \begin{math}Y_i\end{math} and \begin{math}Z_i\end{math}), and by \nset{}
    the cardinality of the sets \begin{math}S_i\end{math}.  Moreover, for all \begin{math}i \in \{1,
            \dots, t\}\end{math} we define \begin{math}X_i = \{x_{i, 1}, \dots, x_{i, n}\}, Y_i = \{y_{i,
        1}, \dots, y_{i, n}\}, Z_i = \{z_{i, 1}, \dots, z_{i, n}\}\end{math}, and \begin{math}S_i =
    \{s_{i, 1}, \dots, s_{i, m}\}\end{math}. We also assume that \begin{math}t = 2^q\end{math} for some
    \begin{math}q \in \mathbb{N}\end{math} (if it is not the case, we add a sufficiently number of dummy
    yes-instances).

    In the following we construct three sets \begin{math}(X^*, Y^*, Z^*)\end{math} of
    \begin{math}nt\end{math} vectors each: \begin{math}X^* = \{x^*_{i, j}\}_{i=1, \dots, t}^{j=1,
            \dots, n}\end{math}, \begin{math}Y^* = \{y^*_{i, j}\}_{i=1, \dots, t}^{j=1, \dots,
            n}\end{math} and \begin{math}Z^* = \{z^*_{i, j}\}_{i=1, \dots, t}^{j=1, \dots,
    n}\end{math}, where each vector is composed of \begin{math}p^*=m+2mq\end{math}
    components. Let us first describe the first \nset{} components of each vector. For all
    \begin{math}i \in \{1, \dots, t\}\end{math}, \begin{math}j \in \{1, \dots,
    n\}\end{math} and \begin{math}k \in \{1, \dots, m\}\end{math} we set:

    \begin{displaymath}
        x^*_{i, j}[k] = \left\{
            \begin{array}{cl}
                1 & \mbox{ if the hyperedge } s_{i, k} \mbox{ contains } x_{i, j} \\ 
                0 & \mbox { otherwise}
            \end{array}
        \right.
    \end{displaymath}
    \begin{displaymath}
        y^*_{i, j}[k] = \left\{
            \begin{array}{cl}
                1 & \mbox{ if the hyperedge } s_{i, k} \mbox{ contains } y_{i, j} \\ 
                0 & \mbox { otherwise}
            \end{array}
        \right.
    \end{displaymath}
    \begin{displaymath}
        z^*_{i, j}[k] = \left\{
            \begin{array}{cl}
                1 & \mbox{ if the hyperedge } s_{i, k} \mbox{ contains } z_{i, j} \\ 
                0 & \mbox { otherwise}
            \end{array}
        \right.
    \end{displaymath}

    Then, for all \begin{math}i \in \{1, \dots, t\}\end{math}, we append two vectors
    \begin{math}b_i\end{math} and \begin{math}\bar{b_i}\end{math} to all vectors
    \begin{math}\{x^*_{i, j}\}_{j=1, \dots, n}\end{math}, \begin{math}\{y^*_{i, j}\}_{j=1, \dots,
    n}\end{math} and \begin{math}\{z^*_{i, j}\}_{j=1, \dots, n}\end{math}. The vector
    \begin{math}b_i\end{math} is composed of \begin{math}mq\end{math} coordinates, and is defined as
    the binary representation of the integer \begin{math}i\end{math}, where each bit is duplicated
    \nset{} times. Finally, \begin{math}\bar{b_i}\end{math} is obtained by taking the complement of
    \begin{math}b_i\end{math} (\textit{i.e.} replacing all zeros by ones, and conversely) as
    depicted in Figure~\ref{fig:cross_compo}. It is now clear that each vector \begin{math}x^*_{i,
    j}\end{math} (resp. \begin{math}y^*_{i, j}\end{math}, \begin{math}z^*_{i, j}\end{math}) is
    composed of \begin{math}p^* = m + 2mq\end{math} coordinates. Thus, the parameter of the input
    instance is a polynomial in \begin{math}n, m\end{math} and \begin{math}\log t\end{math} whereas
    the total size of the instance is a polynomial in the size of the sequence of inputs, as
    required in cross-compositions.  It now remains to prove that \begin{math}(X^*, Y^*,
    Z^*)\end{math} contains an assignment of cost \begin{math}k^* = nt (mq + m -1)\end{math} if and
    only if for all \begin{math}i \in \{1, \dots, t\}\end{math}, \begin{math}S_i\end{math} contains
    a perfect matching \begin{math}S'_i\end{math}.

    \begin{itemize}
	    \item Suppose that for all \begin{math}i \in \{1, \dots, t\}\end{math} we have a perfect
            matching \begin{math}S'_i \subseteq S_i\end{math}. W.l.o.g. suppose that
            \begin{math}S'_i = \{s_{i, 1}, \dots, s_{i, n}\}\end{math}. Then, for each
            \begin{math}j \in \{1, \dots, n\}\end{math}, we have \begin{math}s_{i, j} = (x_{i,
                j_1}, y_{i, j_2}, z_{i, j_3})\end{math} for some \begin{math}j_1, j_2, j_3 \in
            \{1, \dots, n\}\end{math}. We assign \begin{math}x^*_{i, j_1}\end{math} with
            \begin{math}y^*_{i, j_2}\end{math} and \begin{math}z^*_{i, j_3}\end{math}. It is
            easy to see that the cost of this triple is \begin{math}m-1+mq\end{math}. Indeed,
            they all have a one at the \begin{math}j^{th}\end{math} coordinate, corresponding to
            the \begin{math}j^{th}\end{math} hyperedge of \begin{math}S_i\end{math} (and this is
                the only shared one, since we can suppose that all hyperedges are pairwise
            distinct), and they all contain the same vectors \begin{math}b_i\end{math} and
            \begin{math}\bar{b_i}\end{math}. Summing up for all instances, we get the desired
            solution value.
	    \item Conversely, first remark that in any assignment, the cost of every triple
            \begin{math}(x^*_{i_1, j_1}, y^*_{i_2, j_2}, z^*_{i_3, j_3})\end{math} is at least
            \begin{math}m-1+mq\end{math}, and let us prove that this bound is tight when (1) all
            elements are chosen within the same instance, \textit{i.e.} \begin{math}i_1 = i_2 = i_3
            = i\end{math}, and (2) this triple corresponds to an element of
            \begin{math}S_i\end{math}, \textit{i.e.} \begin{math}(x_{i, j_1}, y_{i, j_2}, z_{i,
            j_3}) \in S_i\end{math}. Indeed, suppose first that \begin{math}i_1 \neq i_2\end{math}.
            Then, since the binary representation of \begin{math}i_1\end{math} and
            \begin{math}i_2\end{math} differs on at least one bit, it is clear that the resulting
            vector is of cost at least \begin{math}m(q+1) > mq + m-1\end{math}. Now if
            \begin{math}i_1 = i_2 = i_3 = i\end{math}, then the result is straightforward, since at
            most one hyperedge of \begin{math}S_i\end{math} can contain \begin{math}x^*_{i_1,
                    j_1}\end{math}, \begin{math}y^*_{i_2, j_2}\end{math} and \begin{math}z^*_{i_3,
            j_3}\end{math}. Finally, using the same arguments as previously, we can easily deduce a perfect
            matching \begin{math}S'_i \subseteq S_i\end{math} for each \begin{math}i \in \{1, \dots,
            t\}\end{math}, and the result follows.
    \end{itemize}
\end{proof}

\def\xscale{2.5}
\def\yscale{1.1}
\def\ysep{0.2}
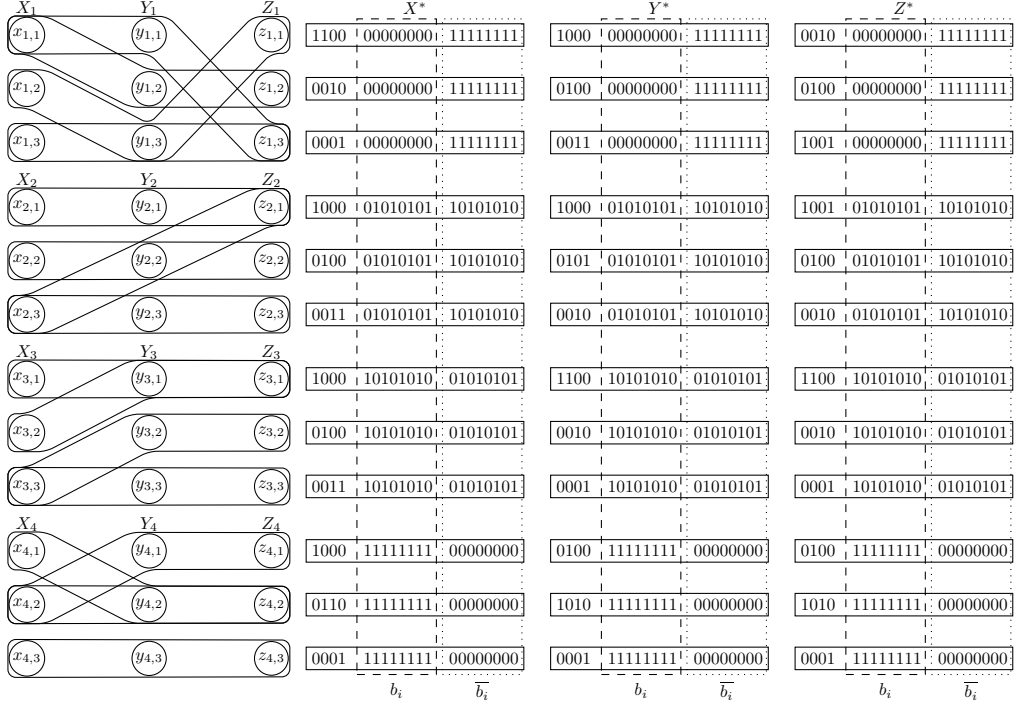
\begin{figure}
    \centering
    \begin{tikzpicture}[%
            nod/.style={circle,draw=black,minimum width=0.5cm, inner sep=1pt},%
            wafer/.style={rectangle, draw=black, minimum width=0.8cm},%
            every node/.style={transform shape},%
			hyperedge/.style={rounded corners=40pt, black},%
            rcorner/.style={rounded corners=3pt},%
            scale=0.65,
            %scale=0.58,%
        ]

        \foreach \yl in {1,2,3,4}{
            \pgfmathparse{(\yl - 1) * (-\ysep - 3) * \yscale} \let\ypos\pgfmathresult
            \foreach \x/\lab in {1/x,2/y,3/z}{
                \foreach \y in {1,2,3}{
                    \pgfmathparse{\yscale * (4 - \y) + \ypos} \let\yp\pgfmathresult
                    \node[nod] (\lab\yl\y) at (\xscale * \x,\yp) {$\lab_{\yl,\y}$};
                    \node[fit=(\lab\yl\y), inner sep=4pt] (f\lab\yl\y) {};
                }
            }
            \foreach \x/\lab in {1/X,2/Y,3/Z}{
                \node (\lab) at (\x * \xscale, \yscale * 3.5 + \ypos) {$\lab_{\yl}$};
            }
        }

        %%%%% Premiere instance
        % <<< Hyperedges: (113), (122), (231), (333)
        \draw[rcorner] 
           (fx11.north east) 
        -- (fy12.north) 
        -- (fz12.north east) 
        -- (fz12.south east) 
        -- (fy12.south west) 
        -- (fx11.south)
        -- (fx11.south west)
        -- (fx11.north west)
        -- cycle;

        \draw[rcorner] 
           (fx11.south east)
        -- (fy11.south)
        -- (fz13.south west)
        -- (fz13.south east)
        -- (fz13.north east)
        -- (fz13.north)
        -- (fy11.north east)
        -- (fx11.north west)
        -- (fx11.south west)
        -- cycle;

        \draw[rcorner] (fx12.south)
        -- (fy13.south west)
        -- (fy13.south east)
        -- (fz11.south)
        -- (fz11.south east)
        -- (fz11.north east)
        -- (fz11.north west)
        -- (fy13.north)
        -- (fx12.north east)
        -- (fx12.north west)
        -- (fx12.south west)
        -- cycle;

        \draw[rcorner] (fx13.north west)
        -- (fz13.north east)
        -- (fz13.south east)
        -- (fx13.south west)
        --cycle;
        % >>>
        % <<< Node labels
        \foreach \x/\y/\lab in {
            1/1/1100\quad00000000\quad11111111,
            1/2/0010\quad00000000\quad11111111,
            1/3/0001\quad00000000\quad11111111,
            3/1/1000\quad00000000\quad11111111,
            3/2/0100\quad00000000\quad11111111,
            3/3/0011\quad00000000\quad11111111,
            5/1/0010\quad00000000\quad11111111,
            5/2/0100\quad00000000\quad11111111,
            5/3/1001\quad00000000\quad11111111
        }{
            \pgfmathparse{(\x + 1.27 * \xscale) * \xscale/1} \let\xpos\pgfmathresult
            \pgfmathparse{(4 - \y) * \yscale} \let\ypos\pgfmathresult
            \node[wafer] (v\x\y) at (\xpos, \ypos) {$\lab$};
        }
        % >>>

        %%%%% Deuxième instance
        % <<< Hyperedges: (111), (222), (333), (321)
        \draw[rcorner] 
           (fx21.north west)
        -- (fz21.north east)
        -- (fz21.south east)
        -- (fx21.south west)
        --cycle;

        \draw[rcorner] 
           (fx22.north west)
        -- (fz22.north east)
        -- (fz22.south east)
        -- (fx22.south west)
        --cycle;

        \draw[rcorner] 
           (fx23.north west)
        -- (fz23.north east)
        -- (fz23.south east)
        -- (fx23.south west)
        --cycle;

        \draw[rcorner]
           (fx23.north)
        % -- (fy22.north west)
        % -- (fy22.north east)
        -- (fz21.north west)
        -- (fz21.north east)
        -- (fz21.south east)
        -- (fz21.south)
        % -- (fy22.south east)
        % -- (fy22.south west)
        -- (fx23.south east)
        -- (fx23.south west)
        -- (fx23.north west)
        -- cycle;
        % >>>
        % <<< Node labels
        \foreach \x/\y/\lab in {
            1/1/1000\quad01010101\quad10101010,
            1/2/0100\quad01010101\quad10101010,
            1/3/0011\quad01010101\quad10101010,
            3/1/1000\quad01010101\quad10101010,
            3/2/0101\quad01010101\quad10101010,
            3/3/0010\quad01010101\quad10101010,
            5/1/1001\quad01010101\quad10101010,
            5/2/0100\quad01010101\quad10101010,
            5/3/0010\quad01010101\quad10101010
        }{
            \pgfmathparse{(\x + 1.27 * \xscale) * \xscale/1} \let\xpos\pgfmathresult
            \pgfmathparse{(4 - \y) * \yscale + (-\ysep - 3) * \yscale} \let\ypos\pgfmathresult
            \node[wafer] (v\x\y) at (\xpos, \ypos) {$\lab$};
        }
        % >>>

        %%%%% Troisième instance
        % <<< Hyperedges: (111), (211), (322), (333)
        \draw[rcorner] 
           (fx31.north west)
        -- (fz31.north east)
        -- (fz31.south east)
        -- (fx31.south west)
        --cycle;

        \draw[rcorner] 
           (fx33.north west)
        -- (fz33.north east)
        -- (fz33.south east)
        -- (fx33.south west)
        --cycle;

        \draw[rcorner]
           (fx32.south east)
        -- (fy31.south)
        -- (fz31.south east)
        -- (fz31.north east)
        -- (fz31.north west)
        -- (fy31.north west)
        -- (fx32.north)
        -- (fx32.north west)
        -- (fx32.south west)
        -- cycle;

        \draw[rcorner]
           (fx33.north)
        -- (fy32.north west)
        -- (fz32.north east)
        -- (fz32.south east)
        -- (fy32.south)
        -- (fx33.south east)
        -- (fx33.south west)
        -- (fx33.north west)
        -- cycle;
        % >>>
        % <<< Node labels
        \foreach \x/\y/\lab in {
            1/1/1000\quad10101010\quad01010101,
            1/2/0100\quad10101010\quad01010101,
            1/3/0011\quad10101010\quad01010101,
            3/1/1100\quad10101010\quad01010101,
            3/2/0010\quad10101010\quad01010101,
            3/3/0001\quad10101010\quad01010101,
            5/1/1100\quad10101010\quad01010101,
            5/2/0010\quad10101010\quad01010101,
            5/3/0001\quad10101010\quad01010101
        }{
            \pgfmathparse{(\x + 1.27 * \xscale) * \xscale/1} \let\xpos\pgfmathresult
            \pgfmathparse{(4 - \y) * \yscale + 2 * (-\ysep - 3) * \yscale} \let\ypos\pgfmathresult
            \node[wafer] (v\x\y) at (\xpos, \ypos) {$\lab$};
        }
        % >>>

        %%%%% Quatrième instance
        % <<< Hyperedges: (122), (211), (222), (333)
        \draw[rcorner] 
           (fx42.north west)
        -- (fz42.north east)
        -- (fz42.south east)
        -- (fx42.south west)
        --cycle;

        \draw[rcorner] 
           (fx43.north west)
        -- (fz43.north east)
        -- (fz43.south east)
        -- (fx43.south west)
        --cycle;

        \draw[rcorner]
           (fx42.south east)
        -- (fy41.south)
        -- (fz41.south east)
        -- (fz41.north east)
        -- (fz41.north west)
        -- (fy41.north west)
        -- (fx42.north)
        -- (fx42.north west)
        -- (fx42.south west)
        -- cycle;

        \draw[rcorner]
           (fx41.north east)
        -- (fy42.north)
        -- (fz42.north east)
        -- (fz42.south east)
        -- (fy42.south west)
        -- (fx41.south)
        -- (fx41.south west)
        -- (fx41.north west)
        -- cycle;
        % >>>
        % <<< Node labels
        \foreach \x/\y/\lab in {
            1/1/1000\quad11111111\quad00000000,
            1/2/0110\quad11111111\quad00000000,
            1/3/0001\quad11111111\quad00000000,
            3/1/0100\quad11111111\quad00000000,
            3/2/1010\quad11111111\quad00000000,
            3/3/0001\quad11111111\quad00000000,
            5/1/0100\quad11111111\quad00000000,
            5/2/1010\quad11111111\quad00000000,
            5/3/0001\quad11111111\quad00000000
        }{
            \pgfmathparse{(\x + 1.27 * \xscale) * \xscale/1} \let\xpos\pgfmathresult
            \pgfmathparse{(4 - \y) * \yscale + 3 * (-\ysep - 3) * \yscale} \let\ypos\pgfmathresult
            \node[wafer] (v\x\y) at (\xpos, \ypos) {$\lab$};
        }
        % >>>

        % <<< Set labels
        \foreach \x/\lab in {1/X,3/Y,5/Z}{
            \pgfmathparse{(\x + 1.27 * \xscale) * \xscale/1} \let\xpos\pgfmathresult
            \pgfmathparse{3.5 * \yscale} \let\ypos\pgfmathresult
            \node (V\x) at (\xpos, \ypos) {$\lab^{*}$};
        }
        % >>>

        % <<< b_t
        \foreach \x in {1,3,5}{
            \pgfmathparse{(\x + 1.08 * \xscale) * \xscale/1} \let\xup\pgfmathresult
            \pgfmathparse{(\x + 1.34 * \xscale) * \xscale/1} \let\xdown\pgfmathresult
            \pgfmathparse{(\xup + \xdown) / 2} \let\xmoy\pgfmathresult

            \pgfmathparse{3.3 * \yscale} \let\yup\pgfmathresult
            \pgfmathparse{0.7 * \yscale + 3 * (-\ysep - 3) * \yscale} \let\ydown\pgfmathresult
            \pgfmathparse{\ydown - 0.3 * \yscale} \let\ymoy\pgfmathresult

            \draw[dashed] (\xup, \yup) rectangle (\xdown, \ydown);
            \node at (\xmoy, \ymoy) {$b_i$};

            \pgfmathparse{(\x + 1.36 * \xscale) * \xscale/1} \let\xup\pgfmathresult
            \pgfmathparse{(\x + 1.62 * \xscale) * \xscale/1} \let\xdown\pgfmathresult
            \pgfmathparse{(\xup + \xdown) / 2} \let\xmoy\pgfmathresult

            \draw[dotted] (\xup, \yup) rectangle (\xdown, \ydown);
            \node at (\xmoy, \ymoy) {$\overline{b_i}$};
        }
		% >>>

    \end{tikzpicture}
    \caption{Example of Cross Composition construction from four equivalent instances of 3-Dimensional
        Perfect Matching
        $(X_1, Y_1, Z_1, \left\{ 
                (x_{1,1}, y_{1,1}, z_{1,3}), 
                (x_{1,1}, y_{1,2}, z_{1,2}), 
                (x_{1,2}, y_{1,3}, z_{1,3}), 
                (x_{1,3}, y_{1,3}, z_{1,3}) 
        \right\})$,
        $(X_2, Y_2, Z_2, \left\{ 
                (x_{2,1}, y_{2,1}, z_{2,1}), 
                (x_{2,2}, y_{2,2}, z_{2,2}), 
                (x_{2,3}, y_{2,3}, z_{2,3}), 
                (x_{2,3}, y_{2,2}, z_{2,1}) 
        \right\})$,
        $(X_3, Y_3, Z_3, \left\{ 
                (x_{3,1}, y_{3,1}, z_{3,1}), 
                (x_{3,2}, y_{3,1}, z_{3,1}), 
                (x_{3,3}, y_{3,2}, z_{3,2}), 
                (x_{3,3}, y_{3,3}, z_{3,3}) 
        \right\})$,
        $(X_4, Y_4, Z_4, \left\{ 
                (x_{4,1}, y_{4,2}, z_{4,2}), 
                (x_{4,2}, y_{4,1}, z_{4,1}), 
                (x_{4,2}, y_{4,2}, z_{4,2}), 
                (x_{4,3}, y_{4,3}, z_{4,3}) 
        \right\})$
        into an instance of bMVA with $\nset = 3,\ \nvec = 12$ and $\ndie = 20$.
    }
    \label{fig:cross_compo}
\end{figure}

\section{Parameterizing according to \texorpdfstring{$\aboveB$}{zeta\_B}}\label{sec:b}

In this section, we present an \begin{math}\FPT{}\end{math} algorithm when parameterized by
\begin{math}\aboveB\end{math} and \nvec{} (recall that both \begin{math}\aboveB\end{math} and
    \nvec{} are smaller parameters than the standard one \begin{math}k\end{math}, since
\begin{math}k = \B+\aboveB\end{math} and \begin{math}n \le k\end{math} in any reduced instance).
Notice first that it is easy to get an \begin{math}\O^*(2^{\aboveB (log(n) +log(p))})\end{math}
algorithm. Indeed, by considering a set \begin{math}i \in [m]\end{math} where \begin{math}c(V^i)
= \B\end{math}, and guessing the positions of the \begin{math}\aboveB\end{math} new zeros (among
\begin{math}np\end{math} possible positions) that will appear in an optimal solution, we can
actually guess in \begin{math}\O^*((np)^{\aboveB})\end{math} the vectors
\begin{math}\{v_{s_j^*}\}\end{math} of an optimal solution, and it remains to check in
polynomial time that every \begin{math}V^j\end{math} can be ``matched'' to
\begin{math}\{v_{s_j^*}\}\end{math}. Now we show how to get rid of the
\begin{math}\log(p)\end{math} term in the exponent.

\begin{nthrm}\label{thm:fptaboveB}
    \probFPTshort can be solved in \begin{math}\O^*(4^{\aboveB \log(n)})\end{math}.
\end{nthrm}

\begin{proof}
    Let \begin{math}\I[m, n, p, k]\end{math} be an instance of our problem and, w.l.o.g., suppose
    that \begin{math}V^1\end{math} is a set whose number of zeros reaches the upper bound
    \begin{math}\B\end{math}, \ie \begin{math}c(V^1) = \B\end{math}. The algorithm consists in
    constructing a solution by finding an optimal assignment between \begin{math}V^1\end{math} and
    \begin{math}V^2, \dots, V^m\end{math}, successively. 

    We first claim that we can decide in polynomial time whether there is an assignment between
    \begin{math}V^1\end{math} and \begin{math}V^2\end{math} which does not create any additional
    zero.

%, or, more formally, whether there is a set of
%\nvec{} disjoint stacks \begin{math}S = \{s_1, \dots, s_n\}\end{math}, where, for all \begin{math}j
%\in [n]\end{math}, each \begin{math}s_j\end{math} is a couple of
%vectors from \begin{math}V^1 \times V^2\end{math}, such that \begin{math}v_{s_j} = v_j^1\end{math}. 

    To that end, we create a bipartite graph \begin{math}G\end{math} with bipartization
    \begin{math}(A, B)\end{math}, \begin{math}A=\{a_1, \dots, a_n\}\end{math},
    \begin{math}B=\{b_1,\allowbreak
    \dots,\allowbreak b_n\}\end{math}, and link \begin{math}a_{j_1}\end{math} and
    \begin{math}b_{j_2}\end{math} for all \begin{math}(j_1, j_2) \in [n] \times [n]\end{math}
    iff assigning vector \begin{math}v_{j_1}^1\end{math} from \begin{math}V^1\end{math} and
    vector \begin{math}v_{j_2}^2\end{math} from \begin{math}V^2\end{math} does not create any
    additional zero in \begin{math}V^1\end{math} (\begin{math}v_{j_1}^1 \wedge v_{j_2}^2 =
    v_{j_1}^1\end{math}). 

%Remark now that the existence of a set of stacks \begin{math}S\end{math} as described previously is
%equivalent to the existence of a perfect matching in \begin{math}G\end{math}. If such a perfect matching can be found,

    If a perfect maching can be found in \begin{math}G\end{math}, then we can safely delete the set
    \begin{math}V^2\end{math} and continue. In order to avoid heavy notations, we consider this
    first step as a polynomial pre-processing, and we re-label \begin{math}V^i\end{math} into
    \begin{math}V^{i-1}\end{math} for all \begin{math}i \in \{3, \dots, m\}\end{math} (and \nset{}
    is implicitly decreased by one).

    In the following, we suppose that the previous pre-processing step cannot apply (\ie there is no
    perfect matching in \begin{math}G\end{math}). Intuitively, in this case any assignment
    (including an optimal one) between \begin{math}V^1\end{math} and \begin{math}V^2\end{math} must
    lead to at least one additional zero in \begin{math}V^1\end{math}. In this case, we perform a
    branching to guess one couple of vectors from \begin{math}V^1 \times V^2\end{math} which will
    induce such an additional zero. More formally, we branch on every couple \begin{math}(j_1, j_2)
    \in [n] \times [n]\end{math}, and create a new instance as a copy of \begin{math}\I\end{math} in
    which \begin{math}v_{j_1}^1\end{math} is replaced by \begin{math}v_{j_1}^1 \wedge
    v_{j_2}^2\end{math}. This operation increases \begin{math}c(V^1)\end{math} by at least one, and
    thus \begin{math}\B\end{math} by at least one as well. If we denote by \begin{math}\I'\end{math}
    this new instance, we can check that a solution of cost at most \begin{math}k\end{math} for
    \begin{math}\I'\end{math} will immediately imply a solution of cost at most
    \begin{math}k\end{math} for \begin{math}\I\end{math}, as \begin{math}\I'\end{math} is
    constructed from \begin{math}\I\end{math} by adding some zeros. The converse is also true as one
    assignment we enumerate corresponds to one from an optimal solution. 

    As the value of \begin{math}\B\end{math} in this branching increases by at least one while we
    still look for a solution of cost \begin{math}k\end{math}, this implies that this branching will
    be applied at most \begin{math}\aboveB\end{math} times. Summing up, we have one polynomial
    pre-processing and one branching of size \begin{math}n^2\end{math} which will be applied at most
    \begin{math}\aboveB\end{math} times. The total running time of this algorithm is thus bounded by
    \begin{math}\O^*(4^{\aboveB \log(n)})\end{math}.

\end{proof}

Despite its simplicity, we now show that, when considering each parameter (\nvec{} and
    \begin{math}\aboveB\end{math}) separately, this algorithm is the best we can hope for (whereas the
existence of an \begin{math}\O^*(2^k)\end{math} algorithm is still open). Indeed, we first show in
Theorem~\ref{thm:lowerboundeth} that the linear dependence in \begin{math}\aboveB\end{math} and
\begin{math}log(n)\end{math} in the exponent is necessary (unless ETH fails), and also that we
cannot hope for an \begin{math}\FPT{}\end{math} algorithm parameterized by
\begin{math}\aboveB\end{math} only unless \begin{math}\FPT{}=\W{2}\end{math}
(Theorem~\ref{thm:whardc}).  Finally, as we will see in the next section
(Theorem~\ref{thm:reducfromcoloring}), this result is matched by a
\begin{math}2^{o(\aboveB)}\end{math} lower bound when \begin{math}n \in \mathbb{N}\end{math} is
fixed. We now present a reduction from the \textsc{Hitting Set} problem which produces an instance
of \probFPTshort.% in which parameters \begin{math}\aboveB\end{math} and \nvec{} are preserved. 

\begin{problem}[H]
    \begin{minipage}{\linewidth}
        \begin{tabular}{m{\headwidth}m{\bodywidth}}
            \textbf{Input:} & \nset{} subsets \begin{math}R_1, ..., R_m\end{math} of
            \begin{math}[n]\end{math}, and an integer \begin{math}k\end{math}\\[\arrayscale]
            \textbf{Question:} & Is there a set \begin{math}R\end{math} of \begin{math}k\end{math}
            elements of \begin{math}[n]\end{math} such that \begin{math}R \cap R_i \neq
            \emptyset\end{math} for any \begin{math}i \in [m]\end{math}?
        \end{tabular}
    \end{minipage}
    \caption{Hitting Set}
    \label{pb:hitting_set}
\end{problem}

\begin{nlemma}\label{lemma:reductionhittingset}
    There is a polynomial reduction from \textsc{Hitting Set} to \probFPTshort that given an
    instance composed of \nset{} subsets of \begin{math}[n]\end{math} and an integer
    \begin{math}k\end{math}, constructs an instance of \probFPTshort \begin{math}\mathcal{I}[m', n',
    p', k']\end{math} such that \begin{math}n' = n\end{math} and \begin{math}\aboveB=k\end{math}.
\end{nlemma}

\begin{proof}
    Let \begin{math}R_1, ..., R_m\end{math} be subsets of \begin{math}[n]\end{math}, and
    \begin{math}k \in \mathbb{N}\end{math}. We construct \nset{} sets \begin{math}V^1, ...,
    V^m\end{math} of \nvec{} vectors each, where, for all \begin{math}i \in [m]\end{math} we have
    \begin{math}V^i = \{v_1^i, ..., v_n^i\}\end{math}, each vector being composed of \nvec{}
    components. For all \begin{math}i \in [m]\end{math} and all \begin{math}j \in [n]\end{math}, if
    \begin{math}j \in R_i\end{math}, then the vector \begin{math}v_j^i\end{math} is composed of ones
    everywhere except at the \begin{math}j^{th}\end{math} component.  If \begin{math}j \notin
    R_i\end{math}, then \begin{math}v_j^i\end{math} is a \begin{math}0\end{math}-vector (\ie a
    vector with zero in every component). We also add a set \begin{math}V^*\end{math} composed of
    \begin{math}(n-1)\end{math} \begin{math}0\end{math}-vectors and one
    \begin{math}1\end{math}-vector as depicted in Figure~\ref{fig:reduction_d_hitting_set}.

    For this constructed instance, it is clear that \begin{math}\B=n(n-1)\end{math} because of the
    set \begin{math}V^*\end{math}. In other words, any assignment will lead to a solution with
    \begin{math}(n-1)\end{math} \begin{math}0\end{math}-vectors, and thus with at least
    \begin{math}n(n-1)\end{math} zeros. We will actually show that this instance has a solution with
    \begin{math}n(n-1)+k\end{math} zeros if and only if \begin{math}R_1, ..., R_m\end{math} has a
    hitting set of size \begin{math}k\end{math}. By the foregoing, we only need to focus on the only
    vector of each set which is assigned to the \begin{math}1\end{math}-vector of
    \begin{math}V^*\end{math}.\\ \begin{math}\Rightarrow\end{math} Let \begin{math}J \subseteq
    [n]\end{math} be a hitting set of size \begin{math}k\end{math}. By the definition of a hitting
    set, for all \begin{math}i \in [m]\end{math}, there exists \begin{math}j_i \in J \cap
    R_i\end{math}. Thus, for all \begin{math}i \in [m]\end{math}, we select the vector
    \begin{math}v_{j_i}^i\end{math} from the set \begin{math}V^i\end{math} to be assigned to the
    \begin{math}1\end{math}-vector of \begin{math}V^*\end{math}. By construction, this vector has
    only one zero at the \begin{math}j_i^{th}\end{math} component, which implies that the
    conjunction of all such vectors \begin{math}\bigwedge_{i=1}^m v_{j_i}^j\end{math} will have a
    \begin{math}1\end{math} everywhere except at the components corresponding to
    \begin{math}J\end{math}. We thus have the desired number of zeros in our solution.\\
    \begin{math}\Leftarrow\end{math} Conversely, for each \begin{math}i \in [m]\end{math}, let
    \begin{math}j_i \in [n]\end{math} be the vector from \begin{math}V^i\end{math} which is assigned
    to the \begin{math}1\end{math}-vector of \begin{math}V^*\end{math}. Since the resulting
    conjunction of all these vectors has only \begin{math}k\end{math} zeros,
    \begin{math}v_{j_i}^i\end{math} cannot be a \begin{math}0\end{math}-vector, and we thus have
    \begin{math}j_i \in R_i\end{math}. Using the same arguments as previously,
    \begin{math}\{u_{j_i}\}_{i \in [m]}\end{math} corresponds to a hitting set of \begin{math}R_1,
    ..., R_m\end{math} of size \begin{math}k\end{math}.\\ As we seen previously,
    \begin{math}\B=n(n-1)\end{math} for the obtained instance, \begin{math}k' = n(n-1)+k\end{math},
    (which implies \begin{math}\aboveB = k\end{math}), and the size of all sets is \nvec{}, as
    desired.
\end{proof}

\def\xscale{0.9}
\def\yscale{0.9}
\def\ysep{0.2}
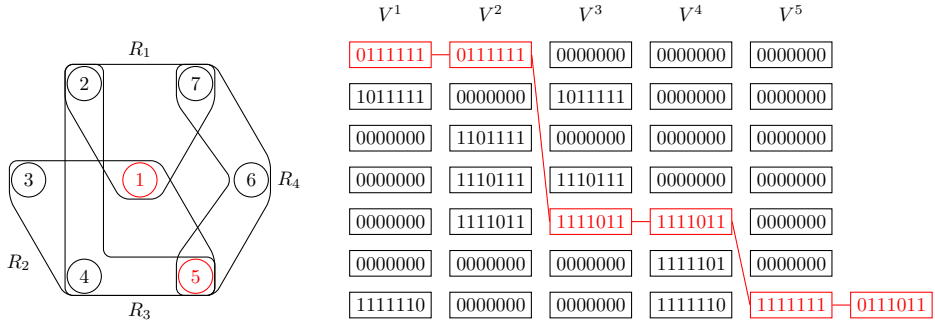
\begin{figure}
    \centering
    \begin{tikzpicture}[%
            nod/.style={circle,draw=black,minimum width=0.5cm},%
            wafer/.style={rectangle, draw=black, minimum width=0.8cm},%
            every node/.style={transform shape},%
			hyperedge/.style={rounded corners=40pt, black},%
            rcorner/.style={rounded corners=3pt},%
            scale=0.74
        ]
        \node[nod,red] (1) at (0,0) {$1$};
        \node[fit=(1), inner sep = 3pt] (f1) {};

        \foreach \cur/\col in {2/black,3/black,4/black,5/red,6/black,7/black}{
            \node[nod, \col] (\cur) at (60 * \cur:2) {$\cur$};
            \node[fit=(\cur), inner sep = 3pt] (f\cur) {};
        }

        \draw[rcorner] (f1.south east)
        -- (f1.south west)
        -- (f2.south west)
        -- (f2.north west)
        -- node[above] {$R_{1}$} (f7.north east)
        -- (f7.south east)
        -- cycle;

        \draw[rcorner] (f1.north)
        -- (f1.north east)
        -- (f5.north east)
        -- (f5.south east)
        -- (f4.south east)
        -- (f4.south west)
        -- node[below left] {$R_{2}$} (f3.south west)
        -- (f3.north west)
        --cycle;

        \draw[rcorner] (f2.north east)
        -- (f4.north east)
        -- (f5.north east)
        -- (f5.south east)
        -- node[below] {$R_{3}$} (f4.south west)
        -- (f2.north west)
        -- cycle;

        \draw[rcorner] (f7.north east)
        -- (f6.north east)
        -- node[right] {$R_{4}$} (f6.south east)
        -- (f5.south east)
        -- (f5.south west)
        -- (f5.north west)
        -- (f6.west)
        -- (f7.south west)
        -- (f7.north west)
        -- cycle;

        \foreach \x/\y/\lab/\id/\col in {
             5/ 2.25/0111111/v11/red,
             5/ 1.50/1011111/v12/black,
             5/ 0.75/0000000/v13/black,
             5/ 0.00/0000000/v14/black,
             5/-0.75/0000000/v15/black,
             5/-1.50/0000000/v16/black,
             5/-2.25/1111110/v17/black,
             7/ 2.25/0111111/v21/red,
             7/ 1.50/0000000/v22/black,
             7/ 0.75/1101111/v23/black,
             7/ 0.00/1110111/v24/black,
             7/-0.75/1111011/v25/black,
             7/-1.50/0000000/v26/black,
             7/-2.25/0000000/v27/black,
             9/ 2.25/0000000/v31/black,
             9/ 1.50/1011111/v32/black,
             9/ 0.75/0000000/v33/black,
             9/ 0.00/1110111/v34/black,
             9/-0.75/1111011/v35/red,
             9/-1.50/0000000/v36/black,
             9/-2.25/0000000/v37/black,
            11/ 2.25/0000000/v41/black,
            11/ 1.50/0000000/v42/black,
            11/ 0.75/0000000/v43/black,
            11/ 0.00/0000000/v44/black,
            11/-0.75/1111011/v45/red,
            11/-1.50/1111101/v46/black,
            11/-2.25/1111110/v47/black,
            13/ 2.25/0000000/v51/black,
            13/ 1.50/0000000/v52/black,
            13/ 0.75/0000000/v53/black,
            13/ 0.00/0000000/v54/black,
            13/-0.75/0000000/v55/black,
            13/-1.50/0000000/v56/black,
            13/-2.25/1111111/v57/red,
            15/-2.25/0111011/s/red%
        }{
            \node[wafer,\col] (\id) at (\x * \xscale, \y) {$\lab$};
        }

        \foreach \x in {1,2,3,4,5}{
            \node (v\x) at (3 * \xscale + \x * 2 * \xscale, 3) {\vset{\x}};
        }

        \draw[red] (v11) -- (v21) -- (v21.east) -- (v35.west) -- (v35) -- (v35.east) -- (v45) --
        (v45.east) -- (v57.west) -- (v57) -- (s);

    \end{tikzpicture}
    \caption{Example of reduction from an instance of Hitting Set consisting of four subsets of
    $[n = 7]: R_1 = \left\{ 1,2,7 \right\}, R_2 = \left\{ 1,3,4,5 \right\}, R_3 = \left\{ 2,4,5
\right\}, R_4 = \left\{ 5,6,7 \right\}$, and an integer $k = 2$, to an instance of \probFPTshort with
$\nset' = 5, \nvec' = n, \ndie = n$.}
    \label{fig:reduction_d_hitting_set}
\end{figure}

As we can see, the reduction is parameter-preserving for \begin{math}\aboveB\end{math}. From the
\begin{math}\W{2}\end{math}-hardness of \textsc{Hitting Set} provided by~\cite{downey2013fundamentals}, we have the
following:

\begin{nthrm}\label{thm:whardc}
    \probFPTshort is \begin{math}\W{2}\end{math}-hard when parameterized by
    \begin{math}\aboveB\end{math}.
\end{nthrm}

As said previously, we also use this reduction to show the following result:

\begin{nthrm}\label{thm:lowerboundeth}
    \probFPTshort cannot be solved in \begin{math}O^*(2^{o(\aboveB) \log(n)})\end{math} nor
    \begin{math}O^*(2^{\aboveB o(\log(n))})\end{math}, unless ETH fails.
\end{nthrm}

\begin{proof}
    To show this, we use a the previous hardness result, but using a constrained version of the
    \textsc{Hitting Set} problem obtained by~\cite{lokshtanov2011slightly}, where the element set is \begin{math}[k]
    \times [k]\end{math} and can thus be seen as a table with \begin{math}k\end{math} rows and
    \begin{math}k\end{math} columns:

    \begin{problem}[H]
        \begin{minipage}{\linewidth}
            \begin{tabular}{m{\headwidth}m{\bodywidth}}
                \textbf{Input:} & An integer \begin{math}k\end{math}, and \begin{math}R_1, ..., R_t
                \subseteq [k] \times [k]\end{math}\\[\arrayscale]
                \textbf{Question:} & Is there a set \begin{math}R\end{math} containing exactly one element from each row such
                that \begin{math}R \cap R_i \neq \emptyset\end{math} for any \begin{math}i \in
                [t]\end{math}?
            \end{tabular}
        \end{minipage}
        \caption{$k \times k$ Hitting Set}
        \label{pb:k_k_hitting_set}
    \end{problem}

    \cite{lokshtanov2011slightly} show that assuming \begin{math}ETH\end{math} this problem cannot be
    solved in \begin{math}2^{o(k \log(k))}n^{O(1)}\end{math} (whereas a simple brute force solves it
    in \begin{math}O^*(2^{k \log(k)})\end{math}).  Notice that we can modify the question of this
    problem by dropping the constraint that \begin{math}S\end{math} contains at least one element
    from each row.  Indeed, let us add to the instance a set of \begin{math}k\end{math} sets
    \begin{math}\{R'_1, \dots, R'_k\}\end{math}, where \begin{math}R'_i\end{math} contains all
    elements of row \begin{math}i\end{math} for \begin{math}i \in [k]\end{math}. Now, finding a
    (classical) hitting set of size \begin{math}k\end{math} on this modified instance is equivalent
    to finding a solution of size \begin{math}k\end{math} for the original instance of
    \textsc{\begin{math}k \times k\end{math}-Hitting Set}.  Moreover, it is easy to check that a
    \begin{math}2^{o(k \log(k))}n^{O(1)}\end{math} algorithm for this relaxed problem would also
    contradict \begin{math}ETH\end{math}.  To summarize, we know that unless
    \begin{math}ETH\end{math} fails, there is no \begin{math}2^{o(k \log(k))}n^{O(1)}\end{math}
    algorithm for the classical \textsc{Hitting Set} problem, even when the ground set has size
    \begin{math}k^2\end{math}.  This allows us to perform the reduction of
    Lemma~\ref{lemma:reductionhittingset} on these special instances, leading to an instance
    \begin{math}\mathcal{I}[m', n', p', k']\end{math} with associated parameter
    \begin{math}\aboveB\end{math} such that \begin{math}\aboveB = k\end{math} and \begin{math}n' =
    k^2\end{math}.  Suppose now that there exists an algorithm for \probFPTshort running in
    \begin{math}2^{o(\aboveB) \log(n)}(k+m+n+p)^{O(1)}\end{math}. Using the reduction above, we
    would be able to solve the instance of \textsc{\begin{math}k \times k\end{math}-Hitting Set} in
    \begin{math}2^{o(k) \log(k^2)}n^{O(1)}\end{math}, and thus in \begin{math}2^{o(k
    \log(k))}n^{O(1)}\end{math}, which would violate \begin{math}ETH\end{math}. A similar idea also
    rules out any algorithm running in \begin{math}2^{\aboveB o(\log(n))}\end{math} under
    \begin{math}ETH\end{math}.
\end{proof}

\section{Parameterizing according to \texorpdfstring{$\abovep$}{zeta\_p}} \label{sec:p}

We now consider the problem parameterized by \begin{math}\abovep = k-p\end{math} (recall that
\begin{math}p \le k\end{math}).  Notice that one motivation of this parameterization is the previous
reduction of Lemma~\ref{lemma:reductionhittingset} from \textsc{Hitting Set}. Indeed, when applied
for \begin{math}n=2\end{math}, it reduces an instance of \textsc{Vertex Cover} to an instance of
\probFPTshort with \begin{math}k=p+\abovep\end{math} where \begin{math}\abovep\end{math} is equal to
the size of the vertex cover. Our intuition is confirmed by the following result: we show that when
parameterized by \begin{math}\abovep\end{math}, the problem is indeed \begin{math}\FPT{}\end{math}
when \begin{math}n=2\end{math} (Theorem~\ref{thm:fptabovep}). We complement this by showing that for
any \begin{math}n \geq 3\end{math}, it becomes \begin{math}\NP{}\end{math}-hard when
\begin{math}\abovep=0\end{math} (Theorem~\ref{thm:reducfromcoloring}), and is thus even not in
\begin{math}\XP{}\end{math}. The reduction we use even proves that for any fixed \begin{math}n \geq
3\end{math}, the problem cannot be solved in \begin{math}2^{o(k)}\end{math} (and thus in
\begin{math}2^{o(\aboveB)}\end{math}) unless \begin{math}ETH\end{math} fails, while the algorithm of
Theorem~\ref{thm:fptaboveB} runs in \begin{math}O^*(2^{O(\aboveB)})\end{math}. In the following,
\nvec{}-\probFPTshort denotes the problem \probFPTshort where the size of all sets is fixed to some
constant \begin{math}n \in \mathbb{N}\end{math}.

\subsection{Positive result for \texorpdfstring{$n=2$}{n = 2}}

In this subsection, we prove that \begin{math}2\end{math}-\probFPTshort is
\begin{math}\FPT{}\end{math} parameterized by \begin{math}\abovep\end{math}. To do so, we reduce to
the \oct problem (\octshort for short). In this problem, given a graph \begin{math}G=(V, E)\end{math}
and an integer \begin{math}c \in \mathbb{N}\end{math}, the objective is to decide whether there exists a partition
\begin{math}(X, S_1, S_2)\end{math} of \begin{math}V\end{math} with \begin{math}|X| \le c\end{math}
such that \begin{math}S_1\end{math} and \begin{math}S_2\end{math} are independent sets.

%\begin{problem}[H]
%\begin{minipage}{\linewidth}
%    \begin{tabular}{m{\headwidth}m{\bodywidth}}
%        \textbf{Input:} & A graph \begin{math}G=(V, E)\end{math}, and integer
%        \begin{math}c\end{math}\\[\arrayscale]
%        \textbf{Question:} & Is there a partition \begin{math}(X, S_1, S_2)\end{math} of
%        \begin{math}V\end{math} such that \begin{math}|X| \le c\end{math} and
%        \begin{math}S_1\end{math} and \begin{math}S_2\end{math} are independent sets ?
%    \end{tabular}
%\end{minipage}
%\caption{\oct}
%\label{pb}
%\end{problem}

We first introduce a generalized version of \octshort, called \bipoct. In this problem, we are given
a set of vertices \begin{math}V\end{math}, an integer \begin{math}c\end{math}, and a set of \nset{}
pairs \begin{math}(A_1, B_1), ..., (A_m, B_m)\end{math} with \begin{math}A_i, B_i \subseteq
    V\end{math} for all \begin{math}i \in [m]\end{math} and \begin{math}A_i \cap B_i =
\emptyset\end{math}. Informally, each pair \begin{math}(A_i, B_i)\end{math} can be seen as a
complete bipartite subgraph. The output of \bipoct is described by a partition \begin{math}(X, S_1,
S_2)\end{math} of \begin{math}V\end{math} such that for any \begin{math}i \in [m]\end{math}, either
(\begin{math}A_i\setminus X \subseteq S_1\end{math} and \begin{math}B_i\setminus X \subseteq
    S_2\end{math}) or (\begin{math}A_i\setminus X \subseteq S_2\end{math} and
\begin{math}B_i\setminus X \subseteq S_1\end{math}). The question is whether there exists
such a partition with \begin{math}|X| \le c\end{math}. As we can see, if all
\begin{math}A_i\end{math} and \begin{math}B_i\end{math} are singletons (and thus form
edges), then \bipoct corresponds to \octshort. Notice that in the following, the considered
parameter of \octshort and \bipoct will always be the standard parameter, \ie
\begin{math}c\end{math}. We first show that there is a linear parameter-preserving reduction
from \begin{math}2\end{math}-\probFPTshort parameterized by \begin{math}\abovep\end{math} to
\bipoct, and then that there is also a linear parameter-preserving transformation from
\bipoct to \octshort.

\begin{nlemma}
    There is a linear parameter-preserving reduction from \begin{math}2\end{math}-\probFPTshort
    parameterized by \begin{math}\abovep\end{math} to \bipoct.
\end{nlemma}

\begin{proof}
    Let \begin{math}\mathcal{I}[m, 2, p, p+\abovep]\end{math} be an instance of
    \begin{math}2\end{math}-\probFPTshort (\ie in which every set contains only two vectors), and
    let us construct an instance \begin{math}\mathcal{I}'\end{math} of \bipoct, such that
    \begin{math}\mathcal{I}\end{math} has a solution of cost \begin{math}p+\abovep\end{math} iff
    \begin{math}\mathcal{I}'\end{math} has a solution of size \begin{math}\abovep\end{math}. Notice
    first that we can suppose that for any \begin{math}i \in [m]\end{math} and any \begin{math}r \in
        [p]\end{math}, we cannot have both \begin{math}v^i_1[r] = 0\end{math} and \begin{math}v^i_2[r] =
    0\end{math} as otherwise any stack \begin{math}s\end{math} from any solution would have
    \begin{math}v_s[r]=0\end{math}, and thus we could safely remove such a component
    \begin{math}r\end{math} from the instance (and decrease \begin{math}k\end{math} and \ndie{} by one).

    Let the vertex set of \begin{math}\mathcal{I}'\end{math} be \begin{math}[p]\end{math}. Then, for
    all \begin{math}i \in [m]\end{math}, let us define \begin{math}A_i = \{r |
    v^i_1[r]=0\}\end{math}, and \begin{math}B_i = \{r | v^i_2[r]=0\}\end{math} as depicted in
    Figure~\ref{fig:min_sum_0_bip_oct}.  By the foregoing,
    and as required in an instance of bip-OCT, we have \begin{math}A_i \cap B_i =
    \emptyset\end{math}.  Let us prove that \begin{math}\mathcal{I}\end{math} has a solution of cost
    \begin{math}p+\abovep\end{math} iff \begin{math}\mathcal{I}'\end{math} has a solution
    \begin{math}(X, S_1, S_2)\end{math} with \begin{math}|X| \le \abovep\end{math}.

    \begin{math}\Rightarrow\end{math}
    Let \begin{math}S=\{s_1,s_2\}\end{math} be a solution of \begin{math}\mathcal{I}\end{math} of
    cost \begin{math}p+\abovep\end{math}.  Let \begin{math}X = \{r |
            v_{s_1}[r]=v_{s_2}[r]=0\}\end{math}, \begin{math}S_1 = \{r | v_{s_1}[r]=0\mbox{ and }
            v_{s_2}[r]=1\}\end{math}, and  \begin{math}S_2 = \{r | v_{s_1}[r]=1\mbox{ and }
    v_{s_2}[r]=0\}\end{math}.  Notice that \begin{math}(X,S_1,S_2)\end{math} forms a partition of
    \begin{math}[p]\end{math} (as we cannot have a \begin{math}r_0\end{math} with
        \begin{math}v_{s_1}[r_0]=v_{s_2}[r_0]=1\end{math}, as this would imply that all the
        \begin{math}nm\end{math} vectors have \begin{math}v[r_0]=1\end{math}, and such cooordinates
    have been removed from the instance in Lemma~\ref{lemma:boundp}), and \begin{math}|X| =
    \abovep\end{math}.  It remains to prove that \begin{math}(X,S_1,S_2)\end{math} is a feasible
    solution of \begin{math}\mathcal{I}'\end{math}.  Let \begin{math}i \in [m]\end{math}.
    Without loss of generality, let us suppose that \begin{math}v^i_1\end{math} has been added
    to \begin{math}s_1\end{math} and \begin{math}v^i_2\end{math} has been added to
    \begin{math}s_2\end{math}.  Let \begin{math}r \in A_i \setminus X\end{math}. Since
    \begin{math}r \in A_i\end{math}, we have \begin{math}v^i_1[r]=0\end{math}, and thus
    \begin{math}v_{s_1}[r]=0\end{math}. Since \begin{math}r \notin X\end{math}, we have
    \begin{math}v_{s_2}[r]=1\end{math}. Thus, \begin{math}r \in S_1\end{math}, which proves
    \begin{math}A_i \setminus X \subseteq S_1\end{math}. Similarly, we can prove that
    \begin{math}B_i\setminus X \subseteq S_2\end{math}.

    \begin{math}\Leftarrow\end{math}
    Let \begin{math}(X,S_1,S_2)\end{math} be a solution of \begin{math}\mathcal{I}'\end{math} with
    \begin{math}|X| \le \abovep\end{math}.  Let \begin{math}s_1\end{math} be such that
    \begin{math}v_{s_1}[r]=0\end{math} iff \begin{math}r \in X\end{math} or \begin{math}r \in
    S_1\end{math}, and let \begin{math}s_2\end{math} be such that \begin{math}v_{s_2}[r]=0\end{math}
    iff \begin{math}r \in X\end{math} or \begin{math}r \in S_2\end{math}.  It remains to prove that
    the solution \begin{math}S=\{s_1,s_2\}\end{math} is feasible, which immediately implies that its
    cost is \begin{math}p + \abovep\end{math}.  Let \begin{math}i \in [m]\end{math}. Without loss of
    generality, let us suppose that \begin{math}A_i \setminus X \subseteq S_1\end{math} and
    \begin{math}B_i \setminus X \subseteq S_2\end{math}. We now claim that
    \begin{math}v^i_1\end{math} can be assigned to \begin{math}s_1\end{math} and
    \begin{math}v^i_2\end{math} can be assigned to \begin{math}s_2\end{math} without creating any
    new zero. To do so, let us show that for all \begin{math}r \in [p]\end{math}, we have
    \begin{math}v_1^i[r] = 0 \implies v_{s_1}[p]=0\end{math} (resp. \begin{math}v_2^i[r] = 0
    \implies v_{s_2}[p]=0\end{math}). Indeed, let \begin{math}r \in [p]\end{math} such that
    \begin{math}v_1^i[r]=0\end{math}. Then by construction, it means that \begin{math}r \in
        A_i\end{math}. Thus, by definition of the solution \begin{math}(X, S_1,
        S_2)\end{math}, it means that either \begin{math}r \in X\end{math} or \begin{math}r
    \in S_1\end{math}, which implies \begin{math}v_{s_1}[r]=0\end{math} as desired. Similar
    arguments show that \begin{math}v_2^i[r] = 0 \implies v_{s_2}[p]=0\end{math} for all
    \begin{math}r \in [p]\end{math}.
\end{proof}

\def\xscale{2.5}
\def\yscale{1.1}
\def\ysep{0.2}
\begin{figure}
    \centering
    \begin{tikzpicture}[%
            nod/.style={circle,draw=black,minimum width=0.5cm, inner sep=1pt},%
            wafer/.style={rectangle, draw=black, minimum width=0.8cm},%
            every node/.style={transform shape},%
			hyperedge/.style={rounded corners=40pt, black},%
            rcorner/.style={rounded corners=7pt, thick},%
            scale=\figscale,
            %scale=0.58,%
        ]

        \foreach \xl/\yl/\lab in {
            0/1/011101111,
            0/2/110110111,
            1/1/110101110,
            1/2/001111011,
            2/1/111111000,
            2/2/111010111,
            3/1/0010\textcolor{BurntOrange}{0}100\textcolor{BurntOrange}{0},
            3/2/1101\textcolor{BurntOrange}{0}011\textcolor{BurntOrange}{0}
        }
        {
            \pgfmathparse{\xl - 5} \let\x\pgfmathresult
            \pgfmathparse{\yl - 1.5} \let\y\pgfmathresult
            \node[wafer] (v\xl\yl) at (\x * \xscale, \y * \yscale) {$\lab$};
        }

        \draw[-] (v01) -- (v12) -- (v21) -- (v31);
        \draw[-] (v02) -- (v11) -- (v22) -- (v32);

        \foreach \lab [evaluate=\lab as \x using \lab-1]  in {1,...,3}{
            \node (V_\lab) at (\x * \xscale - 5*\xscale, 1.25 * \yscale) {\vset{\lab}};
        }

        \node (S) at (-2 * \xscale, 1.25 * \yscale) {$S$};

        \foreach \lab/\x/\col in {4/1/black, 6/2/black, 3/3/black, 5/4/BurntOrange, 1/5/black, 2/6/black, 7/7/black, 8/8/black}{
            \node[nod, \col] (\lab) at (45 * \x:2*\yscale) {$\lab$};
        }

        \node[nod, BurntOrange] (9) at (0,0) {$9$};

        % <<< V1
        %\draw[rcorner, red]
               %(f3.south)
            %-- (f3.west)
            %-- (f3.north)
            %-- (f6.north)
            %-- (f6.east)
            %-- (f6.south)
            %-- cycle;

        %\draw[rcorner, red]
               %(f1.east)
            %-- (f1.south)
            %-- (f1.west)
            %-- (f5.west)
            %-- (f5.north)
            %-- (f5.east)
            %-- cycle;

        \draw[rcorner, red] 
                (145:1.7*\yscale) 
            arc (145:80:1.7*\yscale)
            --  (80:2.3*\yscale)
            arc (80:145:2.3*\yscale) %node[above left, red, midway] {$A_1$}
            --  cycle;

        \node[red] at (-0.55 * \xscale, 2.3 * \yscale) {$A_1$};

        \draw[rcorner, red] 
                (235:1.7*\yscale) 
            arc (235:170:1.7*\yscale)
            --  (170:2.3*\yscale)
            arc (170:235:2.3*\yscale) % node[below left, red, midway] {$B_1$}
            --  cycle;

        \node[red] at (-1.05 * \xscale, -1.2 * \yscale) {$B_1$};

        \draw[*-*, red, line width=2pt]
                (112.5:1.7*\yscale)
            --  (202.5:1.7*\yscale);

        % >>>
        % <<< V2
        \draw[rcorner, Green] 
                (125:2.3*\yscale)
            arc (125:190:2.3*\yscale) % node[above left, Green, midway] {$A_2$}
            --  (0, -0.5 * \yscale)
            arc (-90:90:0.5*\yscale)
            --  cycle;
            
        \node[Green] at (-1.05 * \xscale, 1.2 * \yscale) {$A_2$};

        \draw[rcorner, Green] 
                (215:1.7*\yscale) 
            arc (215:325:1.7*\yscale)
            --  (325:2.3*\yscale)
            arc (325:215:2.3*\yscale) % node[below, Green, midway] {$B_2$}
            --  cycle;

        \node[Green] at (0.00 * \xscale, -2.5 * \yscale) {$B_2$};

        \draw[*-*, Green, line width=2pt]
                (270:1.7*\yscale)
            --  (-0.85*\yscale, -0.5*\yscale);

        % >>>
        % <<< V3
        \draw[rcorner, blue] 
                (35:1.7*\yscale) 
            arc (35:100:1.7*\yscale)
            --  (100:2.3*\yscale)
            arc (100:35:2.3*\yscale) %node[above right, blue, midway] {$A_3$}
            --  cycle;

        \node[blue] at (0.55 * \xscale, 2.3 * \yscale) {$A_3$};

        \draw[rcorner, blue] 
                (-55:2.3*\yscale)
            	arc (-55:10:2.3*\yscale) %node[right, blue, midway] {$B_3$}
            --  (0, 0.5*\yscale)
            arc (90:270:0.5*\yscale)
            --  cycle;

        \node[blue] at (1.05 * \xscale, -1.2 * \yscale) {$B_3$};

        \draw[*-*, blue, line width=2pt]
                (67.5:1.7*\yscale)
            --  (0.85*\yscale, 0.4*\yscale);
		% >>>

        % \node[red, draw=red, rectangle, rounded corners=3pt] (la) at (4*\yscale, 0) {$A$};
        % \node[red, draw=red, rectangle, rounded corners=3pt] (lb) at (6*\yscale, 0) {$B$};
        % \draw[*-*, red, line width=2pt] (la) to (lb);
        % \node[text width=3cm] at (8*\yscale, 0) {$(A,B)$ defines a complete bipartite graph};

    \end{tikzpicture}
    \caption{Example of reduction from an instance of 2-\probFPTshort, with
        $\nvec = 2, \nset = 3, \ndie = 9$, admitting a solution of cost $\ndie
        + 2$, to an instance of \bipoct, with $|V| = \ndie,\ A_1 = \left\{ 3,6
        \right\}, B_1 = \left\{ 1,5 \right\}, A_2 = \left\{ 2,3,7 \right\}, B_2
        = \left\{ 3,5,9 \right\}, A_3 = \left\{ 4,6 \right\}, B_3 = \left\{
    7,8,9 \right\}$, admitting the partition $\left( \left\{ 5,9 \right\}, \left\{ 3, 4, 6 \right\},
    \left\{ 1, 2, 7, 8 \right\} \right)$ as solution of cost $2$.}
    \label{fig:min_sum_0_bip_oct}
\end{figure}

\begin{nlemma}
    There is a linear parameterized reduction from \bipoct to \octshort.
\end{nlemma}

\begin{proof}
    Let \begin{math}\mathcal{I}=(V,\{A_i,B_i\}_{i \in [m]}, c)\end{math} be an instance of \bipoct.
    Let us construct a graph \begin{math}G'=(V', E')\end{math} which contains an odd cycle
    transversal of size \begin{math}c\end{math} if and only if \begin{math}\mathcal{I}\end{math} has
    a solution of size \begin{math}c\end{math} for \bipoct.
%Let us define an instance \begin{math}I_{OCT}=(V',E')\end{math} such that
%\begin{math}Opt_{bip-OCT}(I_{bip-OCT}) \le \abovep\end{math} iff
%\begin{math}Opt_{OCT}(I_{OCT}) \le \abovep\end{math}.
    Observe first that we cannot simply set \begin{math}V'=V\end{math} and \begin{math}E'=\bigcup_{i
            \in [m], a \in A_i, b \in B_i}\{a,b\}\end{math}.  Indeed, if for example \begin{math}A_1 = \{2,
            \dots, n\}\end{math}, \begin{math}B_1 = \{1\}\end{math}, \begin{math}A_2=\{2, \dots,
    \frac{n}{2}\}\end{math} and \begin{math}B_2=\{\frac{n}{2}+1, \dots, n\}\end{math}, defining
    \begin{math}G'\end{math} as above would lead to an odd cycle transversal of size one, as
    removing only vertex \begin{math}\{1\}\end{math} makes the graph bipartite with bipartization
    \begin{math}(A_2, B_2)\end{math}.  However, this solution is not feasible for \bipoct as
    \begin{math}A_1 \setminus X = A_1\end{math}, and \begin{math}A_1 \not \subseteq A_2\end{math}
    and \begin{math}A_1 \not \subseteq  B_2\end{math}. Intuitively, we have to 
% any  \begin{math}Opt(I_{bip-OCT}) > 1\end{math}.
%In the previous example, the problem is that the solution \begin{math}Opt(I_{OCT})\end{math} cannot be translated
%directly into a solution of \begin{math}I_{bip-OCT}\end{math}. Indeed,
%\begin{math}Opt(I_{OCT})\end{math} adds \begin{math}A_2\end{math} in \begin{math}S_1\end{math} and
%\begin{math}B_2\end{math} in
%\begin{math}S_2\end{math}. This corresponds to split \begin{math}A_1 \setminus X (= A_1)\end{math}
%between \begin{math}S_1\end{math} and \begin{math}S_2\end{math}, which is not a
%feasable solution of bip-OCT. Thus, we have to introduce gadgets that 
    prevent solutions of \begin{math}G'\end{math} from splitting sets \begin{math}A_i \setminus
    X\end{math} (and \begin{math}B_i \setminus X\end{math}) between the two parts of the
    bipartization. To do so, we will construct \begin{math}G'\end{math} as described above, and then
    we "augment" each bipartite graph by adding \begin{math}c+1\end{math} new vertices on each side.
    More formally, we start by setting \begin{math}V'=V\end{math} as said before, and for all
    \begin{math}i \in [m]\end{math}, we create two sets of \begin{math}c+1\end{math} new vertices
    \begin{math}A'_i\end{math}, \begin{math}B'_i\end{math}. We then set \begin{math}E' = \bigcup_{i
    \in [m], a \in A_i \cup A'_i, b \in B_i \cup B'_i}\{a,b\}\end{math}.

    Let us now prove that \begin{math}\mathcal{I}\end{math} contains a solution of size
    \begin{math}c\end{math} for \bipoct if and only if \begin{math}G'\end{math} contains an odd
    cycle transversal of size \begin{math}c\end{math}.
%Let us now prove that \begin{math}Opt_{bip-OCT}(I_{bip-OCT}) \le \abovep\end{math} iff \begin{math}Opt_{OCT}(I_{OCT}) \le
%\abovep\end{math}.

    \begin{math}\Rightarrow\end{math}
    Let \begin{math}(X,S_1,S_2)\end{math} be an optimal solution of
    \begin{math}\mathcal{I}\end{math}.  We define a partial solution
    \begin{math}X',S'_1,S'_2\end{math} of \begin{math}G'\end{math} by setting
    \begin{math}X'=X\end{math} and \begin{math}S'_l=S_l\end{math} for \begin{math}l \in \{1,
        2\}\end{math} (the solution is partial in the sense that it remains to assign vertices of
    \begin{math}A'_i \cup B'_i\end{math}, for all \begin{math}i \in [m]\end{math}).  Let
    \begin{math}i \in [m]\end{math}.  If \begin{math}A_i \setminus X = \emptyset\end{math} and
    \begin{math}B_i \setminus X = \emptyset\end{math}, then we add (arbitrarily)
    \begin{math}A'_i\end{math} to \begin{math}S'_1\end{math} and \begin{math}B'_i\end{math} to
    \begin{math}S'_{2}\end{math}.  Otherwise, if \begin{math}A_i \setminus X \neq
    \emptyset\end{math} and is added to \begin{math}S_l\end{math}, we add \begin{math}A'_i\end{math}
    to \begin{math}S'_l\end{math} and \begin{math}B'_i\end{math} to \begin{math}S'_{l'}\end{math}
    with \begin{math}l, l' \in \{1, 2\}, l' \neq l\end{math}, and if \begin{math}B_i \setminus X
    \neq \emptyset\end{math} and is added to \begin{math}S_l\end{math}, we add
    \begin{math}B'_i\end{math} to \begin{math}S'_l\end{math} and \begin{math}A'_i\end{math} to
    \begin{math}S'_{l'}\end{math} with \begin{math}l, l' \in \{1, 2\}, l' \neq l\end{math}.

    This new solution has the same size (\begin{math}|X'|=|X|\end{math}) and we claim that it is an
    odd cycle transversal of \begin{math}G'\end{math}. Indeed, let us check that any edge
    \begin{math}\{u,v\} \in E'\end{math} such that \begin{math}\{u,v\} \cap X' = \emptyset\end{math}
    is not entirely contained in a \begin{math}S'_l\end{math}. If \begin{math}\{u,v\}\end{math} is
    an edge of a complete bipartite of \begin{math}\mathcal{I}\end{math}, \ie if there exists
    \begin{math}i\end{math} such that \begin{math}u \in A_i\end{math} and \begin{math}v \in
    B_i\end{math}, then by definition of the solution \begin{math}(X, S_1, S_2)\end{math} it is
    straightforward that \begin{math}u\end{math} and \begin{math}v\end{math} are not both in
    \begin{math}S'_1\end{math} nor in \begin{math}S'_2\end{math}. Otherwise, if
    \begin{math}\{u,v\}\end{math} is adjacent to one or two of the new vertices, let
    \begin{math}i\end{math} be such that \begin{math}u \in A'_i\end{math}. If \begin{math}v \in
    B'_i\end{math}, then the solution is valid as \begin{math}A'_i\end{math} and
    \begin{math}B'_i\end{math} are never added to the same set \begin{math}S'_l\end{math},
    \begin{math}l \in \{1, 2\}\end{math}.  Otherwise, we necessarily have \begin{math}v \in
        B_i\end{math}. Let \begin{math}l \in \{1, 2\}\end{math} be such that \begin{math}B_i \setminus
    X\end{math} (which is not empty) has been added to \begin{math}S'_l\end{math}. In this case
    \begin{math}A_i\end{math} (and thus \begin{math}u\end{math}) has been added to
    \begin{math}S'_{l'}\end{math}, with \begin{math}l' \neq l\end{math}.

    \begin{math}\Leftarrow\end{math}
    Let \begin{math}(X',S'_1,S'_2)\end{math} be an optimal solution of \begin{math}G'\end{math}.
    For any \begin{math}i \in [m]\end{math}, let \begin{math}\tilde{A_i} = (A_i \cup A'_i) \setminus
    X'\end{math} and \begin{math}\tilde{B_i} = (B_i \cup B'_i) \setminus X'\end{math}.  A first
    observation is that \begin{math}\tilde{A_i} \neq \emptyset\end{math} and \begin{math}\tilde{B_i}
    \neq \emptyset\end{math} as \begin{math}|A_i \cup A'_i| = |B_i \cup B'_i| > c\end{math} and
    \begin{math}|X'| \le c\end{math}.  A second observation is that for any
    \begin{math}u\end{math} and \begin{math}v \in \tilde{A_i}\end{math}, \begin{math}u\end{math}
    and \begin{math}v\end{math} are in the same set \begin{math}S'_l\end{math} for some
    \begin{math}l \in \{1, 2\}\end{math}.  Indeed, suppose by contradiction that \begin{math}u
        \in S'_1\end{math} and \begin{math}v \in S'_2\end{math}. As \begin{math}\tilde{B_i} \neq
    \emptyset\end{math}, there exists \begin{math}b \in \tilde{B_i}\end{math} and
    \begin{math}l \in \{1, 2\}\end{math} such that \begin{math}b \in S'_l\end{math}.  As all
    the edges of the complete bipartite subgraph on \begin{math}(A_i \cup A'_i, B_i \cup
        B'_i)\end{math} belong to \begin{math}E'\end{math}, we have \begin{math}\{u,b\} \in
    E'\end{math} and \begin{math}\{v,b\} \in E'\end{math}, and thus \begin{math}S'_l\end{math}
    contains both endpoints of an edge of \begin{math}E'\end{math}, which is a contradiction.
    In the same way, we can prove that for any \begin{math}i \in [m]\end{math}, and any
    \begin{math}u\end{math} and \begin{math}v \in \tilde{B_i}\end{math}, \begin{math}u\end{math}
    and \begin{math}v\end{math} are in the same set \begin{math}S'_l\end{math} for some
    \begin{math}l \in \{1, 2\}\end{math}.

    Thus, according to the two previous observations, for any \begin{math}i \in [m]\end{math} we can
    define \begin{math}\lambda_{\tilde{A_i}} \in \{1,2\}\end{math} and
    \begin{math}\lambda_{\tilde{B_i}} \in \{1,2\}\end{math} such that \begin{math}\tilde{A_i}
        \subseteq S'_{\lambda_{\tilde{A_i}}}\end{math} and \begin{math}\tilde{B_i} \subseteq
        S'_{\lambda_{\tilde{B_i}}}\end{math}, with \begin{math}\lambda_{\tilde{A_i}} \neq
    \lambda_{\tilde{B_i}}\end{math}.

    Let us now define \begin{math}X=X' \cap V\end{math}, \begin{math}S_1 = S'_1 \cap V\end{math},
    and \begin{math}S_2 = S'_2 \cap V\end{math}, and check that this is a valid solution of
    \begin{math}\mathcal{I}\end{math}.  Let \begin{math}i \in [m]\end{math}. Observe first that
    \begin{math}A_i \setminus X \subseteq \tilde{A_i}\end{math}, and thus either \begin{math}A_i
        \setminus X = \emptyset\end{math}, or \begin{math}A_i \setminus X \subseteq
        S_{\lambda_{\tilde{A_i}}}\end{math}.  As the same fact also holds for \begin{math}B_i \setminus
    X\end{math}, and as \begin{math}\lambda_{\tilde{A_i}} \neq \lambda_{\tilde{B_i}}\end{math}, the
    constraint (\begin{math}A_i\setminus X \subseteq S_1\end{math} and \begin{math}B_i\setminus X
        \subseteq S_2\end{math}) or (\begin{math}A_i\setminus X \subseteq S_2\end{math} and
    \begin{math}B_i\setminus X \subseteq S_1\end{math}) is respected, and the solution is feasible,
    which concludes the proof.

\end{proof}

As \oct can be solved in \begin{math}O^*(2.3146^c)\end{math}, proved by~\cite{lokshtanov2014faster}, and since our
parameters are exactly preserved in our two reductions, we obtain the following result:
\begin{nthrm}\label{thm:fptabovep}
    \begin{math}2\end{math}-\probFPTshort can be solved in \begin{math}O^*(d^{\abovep})\end{math}
    where \begin{math}d \le 2.3146\end{math} is such that
    \octshort can be solved in \begin{math}O^*(d^c)\end{math}.
\end{nthrm}

\subsection{Negative results for \texorpdfstring{$n \geq 3$}{n <= 3}}

We now complement the previous result by proving that the problem is intractable with respect to the
parameter \begin{math}\abovep\end{math} for larger values of \nvec{}.

\begin{nthrm}\label{thm:reducfromcoloring}
    For any fixed \begin{math}n \geq 3\end{math}, \nvec{}-\probFPTshort is not in
    \begin{math}\XP{}\end{math} when parameterized by \begin{math}\abovep\end{math} (unless
    \begin{math}\P{} = \NP{}\end{math}), and cannot be solved in \begin{math}2^{o(k)}\end{math}
    (unless \begin{math}ETH\end{math} fails). 
\end{nthrm}

\begin{proof}
    Let \begin{math}\chi \geq 3\end{math}. We present a reduction from
    \begin{math}\chi\end{math}-\textsc{Coloring}. Given a graph \begin{math}G=(V,
    E)\end{math}, this problem consists to ask for a mapping \begin{math}f:V \longrightarrow [\chi]\end{math} such
    that for all \begin{math}\{u, v\} \in E\end{math} we have \begin{math}f(u) \neq
        f(v)\end{math}.  Let \begin{math}E = \{e_1, ..., e_{m_G}\}\end{math} and \begin{math}V =
    [n_G]\end{math}. Let us construct an instance \begin{math}\mathcal{I}\end{math} of
    \nvec{}-\probFPTshort with \begin{math}n=\chi\end{math}, \begin{math}p=n_G\end{math},
    \begin{math}m=m_G\end{math} and such that \begin{math}G\end{math} admits a
    \begin{math}\chi\end{math}-coloring iff \begin{math}I\end{math} has a solution of cost
    \ndie{} (\ie \begin{math}\abovep=0\end{math}). To each edge \begin{math}e_i= \{u,v\} \in
    E\end{math}, \begin{math}i \in [m_G]\end{math}, we associate a set \begin{math}V^i\end{math}
    with \begin{math}|V^i|=\chi\end{math}, where:
    \begin{itemize}
	    \item \begin{math}v^i_1\end{math} represents the vertex \begin{math}u\end{math}, that is
            \begin{math}v^i_1[u]=0\end{math} and \begin{math}v^i_1[r]=1\end{math} for any
            \begin{math}r
            \in [n_G]\end{math}, \begin{math}r \neq u\end{math},
	    \item \begin{math}v^i_2\end{math} represents the vertex \begin{math}v\end{math}, that is
            \begin{math}v^i_2[v]=0\end{math} and \begin{math}v^i_2[r]=1\end{math} for any
            \begin{math}r
            \in [n_G]\end{math}, \begin{math}r \neq v\end{math},
	    \item for all \begin{math}j \in \{3, \dots, \chi\}\end{math}, \begin{math}v^i_j\end{math} is
            a \begin{math}1\end{math}-vector, \ie it has a \begin{math}1\end{math} at every
            component.
    \end{itemize} 
    An example of this construcion is depicted in Figure~\ref{fig:k_color}.
    Let us now prove that \begin{math}G\end{math} admits a \begin{math}\chi\end{math}-coloring iff
    \begin{math}\mathcal{I}\end{math} has a solution of cost \begin{math}p=n_G\end{math}.

    \begin{math}\Rightarrow\end{math} 
    Let \begin{math}S_j \subseteq V\end{math}, \begin{math}j \in
    [\chi]\end{math} be the \begin{math}\chi\end{math} color classes (notice that the
        \begin{math}S_j\end{math} are pairwise disjoint, some of them may be empty, and
    \begin{math}\bigcup_{j \in [\chi]} S_j = V\end{math}).  To each
    \begin{math}S_j\end{math} we associate a stack \begin{math}s_j\end{math} such that
    \begin{math}v_{s_j}[r]=0\end{math} iff \begin{math}r \in S_j\end{math}.  It remains to
    prove that the solution \begin{math}S=\{s_1,\dots,s_{\chi}\}\end{math} is feasible, as
    its cost is exactly \ndie{} by construction.  Let us consider a set
    \begin{math}V^i\end{math} where \begin{math}v^i_1\end{math} (resp.
        \begin{math}v^i_2\end{math}) represents a vertex \begin{math}u\end{math} (resp.
    \begin{math}v\end{math}). As \begin{math}\{u,v\}\end{math} is an edge of
    \begin{math}G\end{math}, we know that \begin{math}u\end{math} and \begin{math}v\end{math}
    have two different colors, \ie that \begin{math}u \in S_{j}\end{math} and \begin{math}v \in
        S_{j'}\end{math}, for some \begin{math}j, j' \in [\chi]\end{math} with \begin{math}j \neq
    j'\end{math}. Thus, we can add \begin{math}v^i_1\end{math} to stack \begin{math}s_j\end{math},
    \begin{math}v^i_2\end{math} to stack \begin{math}s_{j'}\end{math}, and the
    \begin{math}\chi-2\end{math} other \begin{math}v^i_j\end{math} (\begin{math}j \geq 3\end{math})
    in an arbitrary way. Since the only \begin{math}0\end{math} in \begin{math}v^i_1\end{math}
    (resp. \begin{math}v^i_2\end{math}) is at the \begin{math}u^{th}\end{math} (resp.
    \begin{math}v^{th}\end{math}) component, we have \begin{math}v^i_1 \wedge v_{s_j} =
    v_{s_j}\end{math} (resp. \begin{math}v^i_2 \wedge v_{s_{j'}} = v_{s_{j'}}\end{math}), which
    proves that \begin{math}S\end{math} is feasible.

    \begin{math}\Leftarrow\end{math}
    Let \begin{math}S=\{s_1,\dots,s_{\chi}\}\end{math} be the stacks of an optimal solution.  For
    \begin{math}j \in [\chi]\end{math}, let \begin{math}S_j = \{r \in [p] |
    v_{s_j}[r]=0\}\end{math}. Notice that \begin{math}\bigcup_{j=1}^{\chi} S_j = V\end{math}, and as
    \begin{math}I\end{math} is of cost \ndie{}, all the \begin{math}S_j\end{math} are pairwise
    disjoints and form a partition of \begin{math}V\end{math}.  Moreover, as for any \begin{math}i
    \in [m]\end{math}, \begin{math}v^i_1\end{math} and \begin{math}v^i_2\end{math} have been
    assigned to different stacks, the corresponding vertices have been assigned to different colors,
    and thus each \begin{math}S_j\end{math} induces an independent set, which completes the
    reduction.

    It is known, thanks to~\cite{impagliazzo2001problems} that there is no \begin{math}2^{o(|V|)}\end{math}
    algorithm for deciding whether a graph \begin{math}G=(V, E)\end{math} admits a
    \begin{math}\chi\end{math}-\textsc{Coloring}, for any \begin{math}\chi \geq 3\end{math} (under
    \begin{math}ETH\end{math}). As we can see, the value of the optimal solution for
    \nvec{}-\probFPTshort in the previous reduction equals the number of vertices in the instance of
    \begin{math}\chi\end{math}-\textsc{Coloring}, which proves that \nvec{}-\probFPTshort cannot be
    solved in \begin{math}2^{o(k)}\end{math} for any \begin{math}n \geq 3\end{math}.

\end{proof}

\def\xscale{1.5}
\def\yscale{0.75}
\def\ysep{0.2}
\begin{figure}
    \centering
    \begin{tikzpicture}[%
            nod/.style={circle,draw=black,minimum width=0.5cm, inner sep=1pt},%
            wafer/.style={rectangle, draw=black, minimum width=0.8cm},%
            every node/.style={transform shape},%
			hyperedge/.style={rounded corners=40pt, black},%
            rcorner/.style={rounded corners=7pt, thick},%
            scale=\figscale,
            %scale=0.58,%
        ]

		% <<< Graph Nodes
		\node[nod, fill=blue!60]  (1) at (-4, 1) {$1$};
		\node[nod, fill=red!70]   (2) at (-3, 0) {$2$};
		\node[nod, fill=Green!60] (3) at (-3, 1) {$3$};
		\node[nod, fill=red!70]   (4) at (-3, 2) {$4$};
		\node[nod, fill=blue!60]  (5) at (-2, 1) {$5$};
		% >>>

		% <<< Graph Edges
		\draw[-] (1) -- (4);
		\draw[-] (1) -- (3);
		\draw[-] (1) -- (2);
		\draw[-] (2) -- (3);
		\draw[-] (3) -- (4);
		\draw[-] (2) -- (5);
		\draw[-] (4) -- (5);
		% >>>

		% <<< Wafers sets
		\foreach \xl/\yl/\lab/\col in {
			0/0/11111/black,
			0/1/10111/black,
			0/2/01111/black,
			1/0/11111/black,
			1/1/11011/black,
			1/2/01111/black,
			2/0/11111/black,
			2/1/11101/black,
			2/2/01111/black,
			3/0/11111/black,
			3/1/11011/black,
			3/2/10111/black,
			4/0/11111/black,
			4/1/11101/black,
			4/2/11011/black,
			5/0/11111/black,
			5/1/11110/black,
			5/2/10111/black,
			6/0/11111/black,
			6/1/11110/black,
			6/2/11101/black,
			7/0/11011/Green!60,
			7/1/01110/blue!60,
			7/2/10101/red!70
		}
		{
			\pgfmathparse{\xl * \xscale} \let\x\pgfmathresult
			\pgfmathparse{\yl * \yscale + 0.25} \let\y\pgfmathresult
			\node[wafer,\col] (v_\xl_\yl) at (\x, \y) {$\lab$};
		}

		\foreach \xl/\yl/\lab in {
			0/3/\vset{1,2},
			1/3/\vset{1,3},
			2/3/\vset{1,4},
			3/3/\vset{2,3},
			4/3/\vset{3,4},
			5/3/\vset{2,5},
			6/3/\vset{4,5},
			7/3/S
		}
		{
			\pgfmathparse{\xl * \xscale} \let\x\pgfmathresult
			\pgfmathparse{\yl * \yscale + 0.25} \let\y\pgfmathresult
			\node (v_\xl) at (\x, \y) {$\lab$};
		}

		\draw[blue]  (v_0_2) -- (v_1_2) -- (v_2_2) -- (v_3_0) -- (v_4_0) -- (v_5_1) -- (v_6_1) -- (v_7_1);
		\draw[red]   (v_0_1) -- (v_1_0) -- (v_2_1) -- (v_3_2) -- (v_4_1) -- (v_5_2) -- (v_6_2) -- (v_7_2);
		\draw[Green] (v_0_0) -- (v_1_1) -- (v_2_0) -- (v_3_1) -- (v_4_2) -- (v_5_0) -- (v_6_0) -- (v_7_0);
	\end{tikzpicture}
	\caption{Example of reduction from a positive instance of \chicol, with $\chi = 3,\ V =
		[5], E = \left\{ \left\{ 1,2 \right\}, \left\{ 1,3 \right\}, \left\{ 1,4
			\right\}, \left\{ 2,3 \right\}, \left\{ 3,4 \right\}, \left\{ 2,5 \right\},
		\left\{ 4,5 \right\} \right\}$, to an instance of \probFPTshort with $\nset = |E| = 6,\
	\nvec = \chi = 3, \ndie = |V| = 5$ admitting a solution of cost $\ndie$.}
	\label{fig:k_color}
\end{figure}

Finally, remark that as for the parameterization by \ndie{}, one could ask if \probFPTshort is
\FPT{} when parameterized by the first lower bound \begin{math}\mathcal{B}\end{math}.  However, we
can see in the previous reduction that we obtain a graph with \begin{math}\mathcal{B}=2\end{math},
and thus the problem is even not in \begin{math}\XP{}\end{math} unless
\begin{math}\P{}=\NP{}\end{math}.

\section{Conclusion}

In this article, we presented some negative and positive results for a multidimensional binary
vector assignment problem in the framework of parameterized complexity. Notice that neither lower
bounds of Theorem~\ref{thm:lowerboundeth} nor Theorem~\ref{thm:reducfromcoloring} are able to rule
out an algorithm running in \begin{math}O^*(2^k)\end{math} (when \nvec{} is part of the input),
hence the existence of such an algorithm seems a challenging open problem. Another interesting
question concerns the improvement of the \begin{math}O(k^2m)\end{math} kernel of
Theorem~\ref{thm:kernelkm} by getting rid of the parameter \nset{}: does \probFPTshort admit a
polynomial kernel when parameterized by \begin{math}k\end{math} only?

\bibliographystyle{apalike}
\bibliography{biblio}

\end{document}